%% file: main.tex
\documentclass[format=acmsmall, review=false, screen=true]{acmart}

\usepackage{graphicx}
\usepackage{array}
\usepackage{url}
\usepackage{fancybox}
\usepackage{multirow}
\usepackage{amssymb}
\usepackage{color}

\usepackage{colortbl}
\usepackage{balance}
\usepackage{fancyhdr}

\usepackage{subfig}
\usepackage{diagbox}
\usepackage{bbding}
\usepackage{placeins}

\usepackage{booktabs}
\usepackage{enumitem}
\usepackage{xcolor,lipsum}
\usepackage[linesnumbered, ruled,vlined]{algorithm2e}

\definecolor{light-gray}{rgb}{.906,  .902,  .902}

\newcommand{\rev}[1]{\textcolor{black}{#1}}
\newcommand{\revv}[1]{\textcolor{black}{#1}}
\newcommand{\minor}[1]{\textcolor{black}{#1}}

\acmJournal{TOSEM}
\acmVolume{9}
\acmNumber{4}
\acmArticle{39}
\acmYear{2019}
\acmMonth{3}
\copyrightyear{2009}
\acmArticleSeq{9}

\setcopyright{acmlicensed}

\acmDOI{0000001.0000001}

\received{Mar 2019}
\received[revised]{?? 2019}
\received[accepted]{?? 2019}

\usepackage{bm}

\begin{document}
\title{Technical Q\&A Site Answer Recommendation via Question Boosting}
\titlenote{Corresponding Authors: Xin Xia}

\author{Zhipeng GAO}
\affiliation{%
  \institution{Monash University}
  \city{Melbourne,}
  \state{VIC}
  \postcode{3168}
  \country{Australia}
  }
\email{zhipeng.gao@monash.edu}

\author{Xin Xia}
\affiliation{%
  \institution{Monash University}
  \city{Melbourne,}
  \state{VIC}
  \postcode{3168}
  \country{Australia}
  }
\email{xin.xia@monash.edu}

\author{David Lo}
\affiliation{%
  \institution{Singapore Management University}
  \city{Singapore,}
    \country{Singapore}
  }
\email{davidlo@smu.edu.sg}

\author{John Grundy}
\affiliation{%
  \institution{Monash University}
  \city{Melbourne,}
  \state{VIC}
  \postcode{3168}
  \country{Australia}
  }
\email{john.grundy@monash.edu}

\begin{abstract}
\input{abstract}

\end{abstract}

%
%

%
%


\maketitle

\renewcommand{\shortauthors}{Zhipeng GAO et al.}

\section{Introduction}
\label{sec:intro}
\input{intro}

\section{Motivation}
\label{sec:pre}
\input{pre}

\section{Our Approach}
\label{sec:approach}
\input{approach}

\section{Automatic Evaluation Experiment Setup}
\label{sec:eval}
\input{eval}

\section{Automatic Evaluation Results}
\label{sec:results}
\input{results}

\section{User Study Setup and Results}
\label{sec:human_results}
\input{human_results}
\section{Discussion}
\label{sec:disc}
\input{disc}

\section{Related Work}
\label{sec:related}
\input{related}

\section{Summary}
\label{sec:con}
\input{conclusion}

\section{Acknowledgements}
\label{sec:ack}
\input{acknowledgements}

\balance
\bibliographystyle{ACM-Reference-Format}
\bibliography{samples}

\end{document}

%% file: abstract.tex
Software developers have heavily used online question and answer platforms to seek help to solve their technical problems. However, a major problem with these technical Q\&A sites is \revv{\emph{"answer hungriness"}} i.e., a large number of questions remain unanswered or unresolved, and users have to wait for a long time or painstakingly go through the provided answers with various levels of quality. 
To alleviate this time-consuming problem, we propose a novel {\sc DeepAns} neural network-based approach to identify the most relevant answer among a set of answer candidates.  Our approach follows a three-stage process: question boosting, label establishment, and answer recommendation. Given a post, we first generate a clarifying question  as a way of question boosting. We automatically establish the \emph{positive}, \emph{neutral$^+$}, \emph{neutral$^-$} and \emph{negative} training samples via label establishment. When it comes to answer recommendation, we sort answer candidates by the matching scores calculated by our neural network-based model. To evaluate the performance of our proposed model, we conducted a large scale evaluation on \revv{four datasets, collected from the real world technical Q\&A sites (i.e., Ask Ubuntu, Super User, Stack Overflow Python and 
Stack Overflow Java). }
Our experimental results show that our approach significantly outperforms several state-of-the-art baselines in automatic evaluation. We also conducted a user study with 50 \revv{solved/unanswered/unresolved questions}. The user study results demonstrate that our approach is effective in solving \revv{the answer hungry problem by recommending the most relevant answers from historical archives.}

%% file: intro.tex
The past decade has witnessed significant social and technical value of Question and Answer (Q\&A) platforms, such as Yahoo! Answers\footnote{https://answers.yahoo.com/}, Quora\footnote{https://www.quora.com/}, and StackExchange\footnote{https://stackexchange.com}. These Q\&A websites have become one of the most important user-generated-content (UGC) portals. 
For example, on the Stack Exchange forums, more than 17 million questions have been asked so far, and more than 11 million pages of these forums are visited daily by users. 
To keep up with the fast-paced software development process, the technical Q\&A platforms have been heavily used by software developers as a popular way to seek information and support via the internet.

\begin{figure}
\centerline{\includegraphics[width=0.80\textwidth]{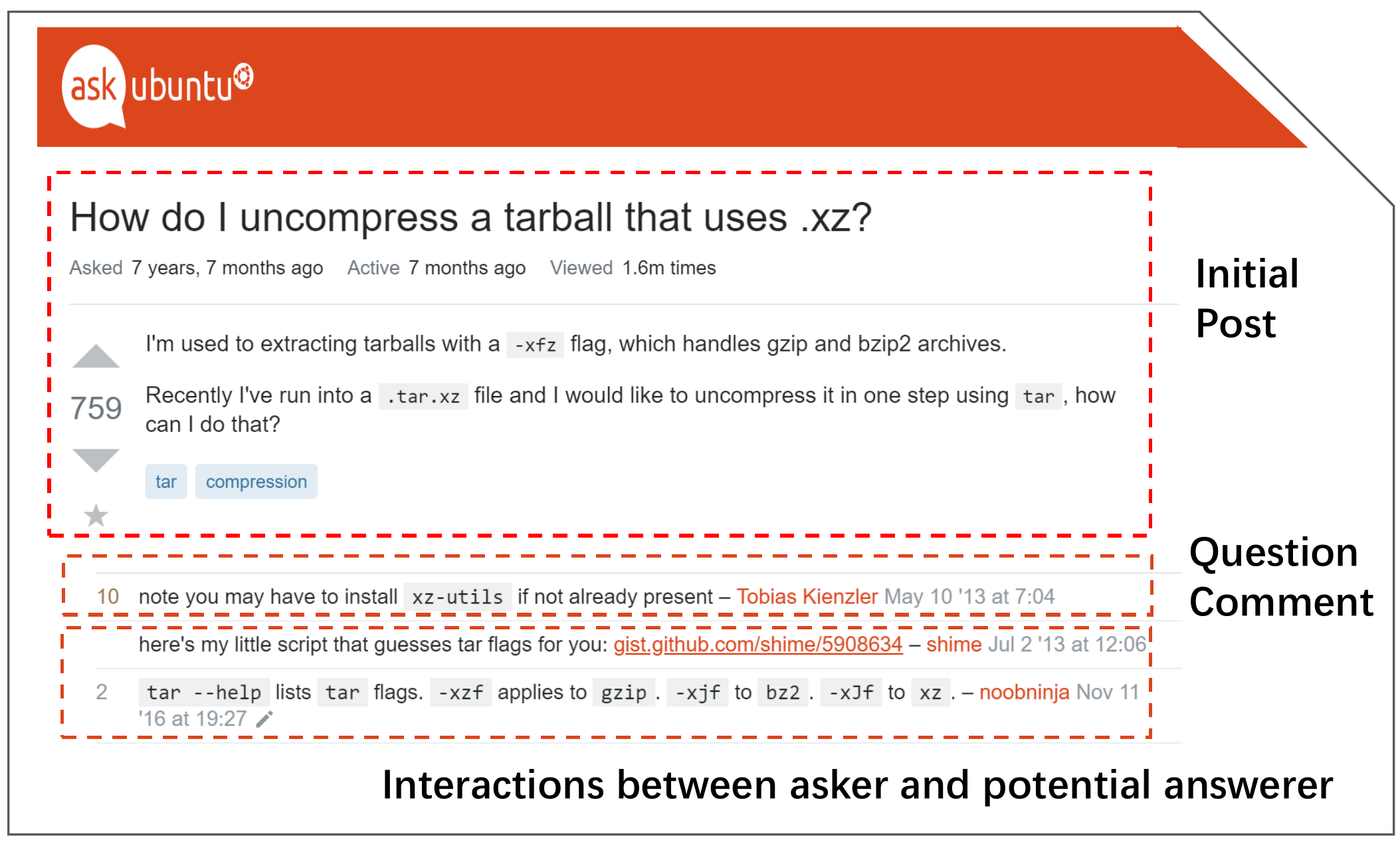}}
\caption{Example Post on Askubuntu}
\label{fig:example}
\end{figure}

StackExchange is a network of online question and answer websites, where each website focuses on a specific topic, such as academia, Ubuntu operating system, latex, etc. There are a lot of technical Q\&A sites which are heavily used by developers, such as Stack Overflow (with a focus on programming-related questions), Ask Ubuntu (with a focus on Ubuntu operating system), Super User (with a focus on computer software and hardware), and Server Fault (with a focus on servers and networks). 
These Q\&A websites allow users to post questions/answers and search for relevant questions and answers. Moreover, if a post is not clear/informative, users routinely provide useful comments to improve the post.
Fig.~\ref{fig:example} shows an example of an initial post and its associated question comment in Ask Ubuntu Q\&A site. By providing the question comment to the original post, it can assist potential helpers to write high quality answers since the question is more informative.


In spite of their success and active user participation, the phenomenon of being \revv{\emph{"answer hungry"}} is still one of the biggest issues within these Q\&A platforms. This concept means that a very large number of questions posted remain \emph{ unanswered and/or unresolved}. 
\revv{
According to our empirical study in different technical Q\&A sites, Ask Ubuntu\footnote{https://askubuntu.com/} and Super User\footnote{https://superuser.com/}, and Stack Overflow\footnote{https://stackoverflow.com/}. 
we found that (1) developers often
have to wait a long time, spanning from days to even many weeks, before getting the first answer to their questions. Moreover, around 20\% of the questions in Ask Ubuntu and Super User  
do not receive any answer at all and leave the askers unsatisfied; and (2) even with provided answers, about 44\% questions in Ask Ubuntu and 39\% questions in Super User are still unresolved, i.e., the question asker does not mark any answer as the accepted solution to their post. In such a case, information seekers have to painstakingly go through the provided answers of various quality with no certainty that a valid answer has been provided.
}

In this work, we aim to address this \revv{answer hungry} phenomenon by recommending the \emph{most relevant} answer or \emph{the best} answer for an unanswered or unresolved question by searching from historical QA pairs. We refer to this problem as \emph{relevant answer recommendation}. We propose a deep learning based approach we name {\sc DeepAns},  which consists of three stages: \emph{question boosting}, \emph{label establishment} and \emph{answer recommendation}.
Given a post, our first step is to generate useful clarifying questions via a trained sequence-to-sequence model. The clarifying question is then appended to the original post as a way of question boosting, which can help eliminate the isolation between question and answers. 
Then, in the label establishment phase, for each enriched question, we pair it with its corresponding answers and automatically label the QA pair as \emph{positive}, \emph{neutral$^+$}, \emph{neutral$^-$} and \emph{negative} samples by leveraging four heuristic rules. 
In the answer recommendation phase, given a question $q$ and an answer candidate $a_i$, our goal is calculating the matching degree of the $\langle$$q$, $a_i$$\rangle$ pair. 
We formulate this problem as a four-category classification problem (i.e., a question and answer pair can be \emph{positive}, \emph{neutral$^+$}, \emph{neutral$^-$}, or \emph{negative} related). We propose a weakly supervised neural network that can be trained with the aforementioned four kinds of training samples.

The key usage scenarios of {\sc DeepAns} are as follows: (1) for unresolved questions which do not have an asker-accepted answer, developers can use {\sc DeepAns} to recommend the best answers; and (2) for  unanswered questions, developers can use {\sc DeepAns} to get the most relevant answers by mining answers to other related questions.

To evaluate the performance of our proposed approach, we conducted comprehensive experiments with \revv{four datasets, collected from the technical Q\&A sites Ask Ubuntu, Super User and Stack Overflow respectively.} 
The large-scale automatic evaluation results suggest that our model outperforms a collection of state-of-the-art baselines by a large margin. For human evaluation, we asked 5 domain experts for their feedback on our generated clarifying questions and answers. Our user study results further demonstrate the effectiveness and superiority of our approach in solving unanswered/unresolved questions. In summary, this paper makes the following contributions:

\begin{itemize}
    \item Previous studies  neglect the value of interactions between the question asker and the potential helper. We argue that a clarifying question between the question and answers is an important aspect of judging the relevance and usefulness of the QA pair. 
    Therefore, we train a sequence-to-sequence model to generate useful clarifying questions for a given post, which can fill the lexical gap between the questions and answers. To the best of our knowledge, this is the first successful application of generating clarifying questions for technical Q\&A sites.
    
    \item We present a novel method to constructing \emph{positive}, \emph{neutral$^+$}, \emph{neutral$^-$}, \emph{negative} training samples via four heuristic rules, which can greatly save the time consuming and labor intensive labeling process.
    
    \item We develop a weakly supervised neural network model for the answer recommendation task. For any question answer pairs, we fit the QA pair into our model to calculate the matching score between them; the higher matching score is estimated by our model, the better chance the answer will be selected as the best answer. In particular, the Q\&A sites can employ our approach as a preliminary step towards marking the potential solution for the unanswered/unresolved question. This can avoid unnecessary time spent by developers to browse questions without an accepted solution.
    
    \item Both our quantitative evaluation and user study show that {\sc DeepAns} can help developers find relevant tehnical question answers more  accurately, compared with state-of-the-art baselines. We have released the source code of  {\sc DeepAns} and the dataset\footnote{\url{https://github.com/beyondacm/DeepAns}} of our study to help other researchers replicate and extend our study.
\end{itemize}

The rest of the paper is organized as follows.
Section~\ref{sec:pre} presents our empirical study of \revv{\emph{answer hungry}} problem in technical Q\&A sites.
Section~\ref{sec:approach} presents the details of our approach to identifying the most relevant answers. 
Section~\ref{sec:eval} presents the experimental set up and evaluation metrics. 
Section~\ref{sec:results} presents the results of our approach on automatic evaluation. 
Section~\ref{sec:human_results} presents the results of our approach on human evaluation. 
Section~\ref{sec:disc} \revv{discusses the strength of our approach and the threats to validity in our study.} 
Section~\ref{sec:related} presents key related work and techniques of this work.
Section~\ref{sec:con} concludes the paper with possible future work.

%% file: pre.tex


\subsection{\revv{Answer Hungry Q\&A Site Phenomenon}}
Despite of -- or perhaps even because of -- the success of technical Q\&A sites, the \emph{\revv{answer hungry}} problem still widely exists in these online forums. We wanted to find out the degree of the problem for technical Q\&A sites. To do this we quantitatively analyzed the prevalence of this \emph{\revv{answer hungry}} issue in 
\revv{real world technical Q\&A sites (i.e., Ask Ubuntu, Super User and Stack Overflow).
Since it is too expensive to run the empirical study on all the Stack Overflow dataset, we only focus on Python and Java related programming language questions in Stack Overflow for our experiment, which refer to \emph{SO (Python)} and \emph{SO (Java)} respectively in this study. The following two metrics are used in our experiment:} 
(1) the proportion of questions that remain unanswered and/or unresolved within these technical Q\&A sites, and (2) the time interval between the posting of one answer and its corresponding question. 

To investigate the proportion of the unanswered and unresolved questions, we first counted the number of questions that have received at least one answer, and refer to these questions as \emph{Answered Questions}. Questions not receiving any answers are referred to as \emph{Unanswered Questions}. For those \emph{Answered Questions}, we further divided them into two groups of \emph{Resolved Questions} and \emph{Unresolved Questions} based on whether any answer within the question thread has been marked or not as the accepted answer by the asker. 
\rev{
Then, we empirically studied the average waiting time measured from the time of question creation to answer posting. We also calculated the average time interval for accepting an answer, which is the time difference between the time a question is created and the time an answer post is accepted.
}
Table~\ref{tab:questionhungrystat} presents the statistical results of our collected data\footnote{For duplicated questions, we only keep the master ones, and remove the others.}. 
\revv{From the table, we have the following observations:
\begin{enumerate}
    \item A large proportion of questions do not receive any answers in these technical Q\&A sites. Consider Ask Ubuntu and Super User as examples -- around 22\% questions in Ask Ubuntu and 19\% questions in Super User do not get any response since the time questions have been created, leaving the askers unsatisfied.
    \item A large amount of questions are still unresolved. For instance, 31.3\% questions in SO (Python) and 35.4\% questions in SO(Java) remain to be unresolved. This phenomenon is probably caused by the following reasons: (a) no good answer was provided within the current question thread, (b) even provided with good answers, it is common for the less experienced users to forget marking a potential answer as a solution.
    \item Developers usually have to wait a long time before getting answers to their questions. It takes on average more than 135 days and 173 days to receive an answer in Ask Ubuntu and Super User sites respectively. The average time to accept an answer is much shorter, which are 18 and 25 days respectively. This further justifies our assumption that users may often forget to mark their accepted answers.
    \item The number of questions posted on Stack Overflow far outnumber the questions posted on Ask Ubuntu and Super User. At the same time, the ratio of the resolved questions in Stack Overflow are also higher than the other two technical Q\&A sites. For instance, 54.5\% questions in SO (Python) were resolved while the same number in Ask Ubuntu was 33.6\%.
    This reflects that, compared with other technical Q\&A sites, Stack Overflow is more popular and more frequently used by the information seekers.  
\end{enumerate}
}

In summary, the \revv{\emph{answer hungry}} phenomenon widely exists and has been one of the biggest challenges in technical Q\&A forums. 

 

\begin{table} 
\caption{\revv{Answer Hungry Statistics}}
\begin{center}
\revv{
\begin{tabular}{||c|l|c||}
    \hline
    \multirow{6}{*}{Ask Ubuntu}   & \# Questions  & 315,924 \\ \cline{2-3}
                              & \# Unanswered Questions  & 69,528 \\ \cline{2-3}
                              & \# Resolved Questions  & 106,301 \\ \cline{2-3}
                              & \# Unresolved Questions & 140,095 \\ \cline{2-3}
                              & Avg Waiting Time & 135.75 (days)   \\ \cline{2-3}
                              & Avg Accepting Time & 18.63 (days)   \\ \cline{2-3}
    \hline
      \multirow{6}{*}{Super User}    & \# Questions  & 380,940 \\ \cline{2-3}
                              & \# Unanswered Questions  & 73,584 \\ \cline{2-3}
                              & \# Resolved Questions & 160,200 \\ \cline{2-3}
                              & \# Unresolved Questions & 147,156 \\ \cline{2-3}
                              & Avg Waiting Time & 173.03 (days)   \\ \cline{2-3}
                              & Avg Accepting Time & 25.69 (days)   \\ \cline{2-3}
    \hline\hline
    \multirow{6}{*}{SO (Python)}   & \# Questions  & 1,236,748 \\ \cline{2-3}
                              & \# Unanswered Questions  & 175,859 \\ \cline{2-3}
                              & \# Resolved Questions  & 674,360 \\ \cline{2-3}
                              & \# Unresolved Questions & 386,529 \\ \cline{2-3}
                              & Avg Waiting Time & 103.97 (days)   \\ \cline{2-3}
                              & Avg Accepting Time & 7.78 (days)   \\ \cline{2-3}
    \hline
      \multirow{6}{*}{SO (Java)}    & \# Questions  & 1,581,814 \\ \cline{2-3}
                              & \# Unanswered Questions  & 213,963 \\ \cline{2-3}
                              & \# Resolved Questions & 808,040 \\ \cline{2-3}
                              & \# Unresolved Questions & 559,811 \\ \cline{2-3}
                              & Avg Waiting Time & 100.52 (days)   \\ \cline{2-3}
                              & Avg Accepting Time & 8.52 (days)   \\ \cline{2-3}
    \hline
\end{tabular}
}
\label{tab:questionhungrystat}
\end{center}
\end{table}

\rev{
\subsection{Clarifying Questions in Technical Q\&A Sites}
\label{subsec:motivation_cqs}
Different from general Q\&A sites, the comments within technical Q\&A sites often include 
\emph{clarifying questions}. 
In technical Q\&A sites, the experts often ask clarifying questions to comments of a post so that they can understand the problem and help the one posting the question. We define a ``clarifying question'' as a question in comments of a post that inquires of missing information for the given post. We wanted to empirically study the proportion and usefulness of clarifying questions in technical Q\&A sites.
}

\revv{
To investigate the proportion of the clarifying questions, we counted the number of comments on the questions as well as the number of comments containing clarifying questions. 
We extracted the clarifying questions as follows: 
We first constructed a Question Comment Set by extracting all the comments on the questions, removing the comments on the answers.
Following that, for each comment in the Question Comment Set, we adopted sentence tokenization method from the NLTK toolkit~\cite{bird2004nltk} to break comment into multiple sentences. We then used the word tokenization method to separate each sentence into a list of tokens and symbols. If the extracted tokens contain the question mark token ``?'', we truncated the sentence till its question mark ``?'' to retrieve the question part of the comment as the clarifying question. If there are multiple clarification questions within the same comment, we kept them as separate clarifying questions.
After that, we removed clarifying questions which are more than 20 words. The results are summarized in Table~\ref{tab:clarifyingquestionstat}.
}

A clarifying question is useful if it helps in getting an answer to a specific question and/or reducing the waiting time.
Imagine a scenario that Bob is a software developer who is seeking help in technical Q\&A sites, he posts a question on these technical Q\&A forums but the question remains unanswered for sometime. Following this, a clarifying question gets asked on the post and then Bob gets an answer. Such a user scenario can help to demonstrate the usefulness of clarifying questions. We estimated the usefulness by calculating the probability of a post getting answered with and without a clarifying question. \revv{The data were collected using the following steps:
\begin{enumerate}
    \item For a given question post, we removed it if the creator of the post responded to his or her own questions. There are around 10\% of the posts being answered by the original question author in these CQA forums. For example, 39,811 and 48,503 questions were removed from the Ask Ubuntu and Super User, while 110,767 and 154,065 questions were removed from the SO (Python) and SO (Java) dataset respectively. 
    \item Considering a question may not have enough time to receive answers if it is posted near the creation date of the data dump, we also removed the unanswered post if it is close (within 7 days) to the release date of the data dump. Since the data dump we used was created on September 5, 2019, we removed the unanswered questions posted after August 25, 2019.
    This results in 521 and 982 unanswered questions were removed from Ask Ubuntu and Super User, while 2,093 and 1,549 unanswered questions were removed from the SO (Python) and SO (Java) dataset respectively. 
    \item For a given clarifying question, we removed it from the candidate list if the clarifying question is posted by the same user of the original question. For such a case, 8\% of the clarifying questions were deleted from Ask Ubuntu and Super User, and 12\% of the clarifying questions were deleted from SO (Python) and SO (Java) respectively. 
    \item Considering a clarification question is helpful only if it was posted before the first answer provided on the thread, we checked the creation date of the clarification question as well as the first answer on the thread, and deleted all the clarification questions posted after the first answers. For example, 12,451 and 18,430 clarification questions were deleted from Ask Ubuntu and Super User, while 54,009 and 82,700 clarification questions were deleted from SO (Python) and SO(Java) dataset respectively.
\end{enumerate}
Finally, we calculated the probabilities of a question receiving answers with and without a clarifying question. The probabilities are defined as follows:
}

\rev{
\begin{equation}
    P\left(\mathbf{A}| \mathbf{CQ} \right) = \frac{  count\left(\mathbf{A}| \mathbf{CQ} \right) }{ count\left(\mathbf{A}| \mathbf{CQ} \right) + 
    count\left(\mathbf{\overline{A}}| \mathbf{CQ} \right) }
\end{equation}
\begin{equation}
    P\left(\mathbf{A}| \mathbf{\overline{CQ}} \right) = \frac{  count\left(\mathbf{A}| \mathbf{\overline{CQ}} \right) }{ count\left(\mathbf{A}| \mathbf{\overline{CQ}} \right) + count\left(\mathbf{\overline{A}}| \mathbf{\overline{CQ}} \right) }
\end{equation}
}

\begin{table} 
\caption{\revv{Clarifying Questions Statistics}}
\revv{
\begin{center}
\begin{tabular}{||c|c|c||}
    \hline
    \multirow{4}{*}{Ask Ubuntu}   & \# Question Comments  & 188,920  \\ \cline{2-3}
                              & \# Clarifying Questions  & 72,359 \\ \cline{2-3}
                              & Pr$\left(\mathbf{A}| \mathbf{CQ} \right)$  & 18.1\% \\ \cline{2-3}
                              & Pr$\left(\mathbf{A}| \mathbf{\overline{CQ}} \right)$ & 14.8\% \\ \cline{2-3}
    \hline
    \multirow{4}{*}{Super User}   & \# Question Comments  & 237,668 \\ \cline{2-3}
                              & \# Clarifying Questions  & 96,296 \\ \cline{2-3}
                              & Pr$\left(\mathbf{A}| \mathbf{CQ} \right)$  & 16.2\% \\ \cline{2-3}
                              & Pr$\left(\mathbf{A}| \mathbf{\overline{CQ}} \right)$ & 12.3\% \\ \cline{2-3}
    \hline\hline  
    \multirow{4}{*}{SO (Python)}   & \# Questions Comments  & 766,490  \\ \cline{2-3}
                              & \# Clarifying Questions  & 329,768 \\ \cline{2-3}
                              & Pr$\left(\mathbf{A}| \mathbf{CQ} \right)$  & 8.4\% \\ \cline{2-3}
                              & Pr$\left(\mathbf{A}| \mathbf{\overline{CQ}} \right)$ & 7.8\% \\ \cline{2-3}
    \hline
    \multirow{4}{*}{SO (Java)}   & \# Questions Comments  & 1,032,176 \\ \cline{2-3}
                              & \# Clarifying Questions  & 467,772 \\ \cline{2-3}
                              & Pr$\left(\mathbf{A}| \mathbf{CQ} \right)$  & 8.1\% \\ \cline{2-3}
                              & Pr$\left(\mathbf{A}| \mathbf{\overline{CQ}} \right)$ & 7.7\% \\ \cline{2-3}
    \hline  
\end{tabular}
\label{tab:clarifyingquestionstat}
\end{center}
}
\end{table}

\rev{
where $\left(\mathbf{A}| \mathbf{CQ} \right)$ and $\left(\mathbf{A}| \mathbf{\overline{CQ}} \right)$ stands for answered posts with and without a clarifying question respectively. Similarly, $\left(\mathbf{\overline{A}}| \mathbf{CQ} \right)$ and $\left(\mathbf{\overline{A}}| \mathbf{\overline{CQ}} \right)$ stands for unanswered posts with and without a clarifying question respectively. The results are summarized in Table~\ref{tab:clarifyingquestionstat}. From the table we have the following observations:
\begin{enumerate}
    \item 
    \revv{In technical Q\&A sites, a large number of comments on questions include clarifying questions. Since our method to extract clarifying questions is rather intuitive, we further sampled 100 clarifying questions from our dataset to do a manual analysis. By manually checking these clarifying questions, we found that 91\% of the clarifying questions are positive clarifying questions.
    The positive clarifying questions often ask more information about the original post, such as ``which version of ubuntu are you using?'', and/or provide potential solutions to the original post, such as ``do you use gnome or kde?''. 
    Only 9\% of the clarifying questions are negative clarifying questions. 
    The negative clarifying questions are often noisy and/or do not appear to provide any useful information for the original post, such as ``did you resolve this?''. These results show that a large proportion of clarifying questions are meaningful and informative.
    }
    \item The likelihood of a post getting an answer with a clarifying question is higher than the likelihood of a post getting an answer without a clarifying question. For example in Ask Ubuntu, without a clarifying question, the probability of a question post to receive answers dropped from 18.1\% to 14.8\%.
    This further justifies our assumption that the clarifying questions are helpful in improving the quality of the original post, hence increasing the chance of a question post receiving  answers. This is why we employ clarifying questions to boost the original question post in our study.
\end{enumerate}
In summary, clarifying questions appear frequently in technical Q\&A sites and can help in improving the original question post and increasing the likelihood of questions to receive answers.
}

In this paper, we aim to invent a new model to not only help developers identify the best answers from a set of candidate answers when they perform QA search activities online, but also recommend the most relevant answers (given to other questions) when they initially post a question online.
Mathematically, let $q$ be the \revv{unanswered or}  unresolved question, let $(a_1, a_2, ..., a_{N})$ be a set of answer candidates,  our task is defined as finding the most relevant answer $a_i^* (i=1, 2, 3, ... N)$, such that:
\begin{equation}
    a^* = argmax_{a_i}P(\mathbf{Accept}|\langle q, a_i\rangle)
\end{equation}
$P(\mathbf{Accept}|\langle q, a_i\rangle)$ 
corresponds to the probability $a_i$ to be accepted given a QA pair $\langle q, a_i\rangle$. 



%% file: approach.tex
We present our approach named {\sc DeepAns}, which ranks candidate answers from a relevant answer pool and recommends the most relevant answer to developers. 
In general, our model follows a three-stage process: \emph{Question Boosting}, \emph{Label Establishment}, and \emph{Answer Recommendation}.
Particularly, in the question boosting phase, {\sc DeepAns} uses an attentional sequence-to-sequence \minor{recurrent neural network} ~\cite{sutskever2014sequence} to generate possible clarifying questions for a given post. These generated questions are appended to the original post as a way of \emph{question boosting}.
Then {\sc DeepAns} automatically constructs \emph{positive}, \emph{neutral$^+$}, \emph{neutral$^-$} and \emph{negative} training samples for each question and answer pair via four heuristic rules.
In the answer recommendation phase, {\sc DeepAns} trains another \minor{convolutional} neural network to calculate the matching score between a given question and a candidate answer, the higher a similarity score is estimated, the more probable the answer will be selected as the best answer. 

\minor{The underlying principle of applying the recurrent networks for the question boosting task is that compared with CNN neural networks, RNN architectures are dedicated sequence models, and this family of architectures has gained tremendous popularity to prominent applications, e.g., machine translation~\cite{bahdanau2014neural, sutskever2014sequence}. For the answer recommendation task, we select the convolutional networks. 
Theoretically, we could also employ the recurrent networks for answer recommendation. However, due to the fact that computing score for each answer in the answer candidate pool is time-consuming, CNN architecture has better performance, lower perplexity, and more importantly, it runs much faster~\cite{dauphin2017language, kim2014convolutional} than RNN architecture for text encoding tasks, i.e., we can process all time steps in parallel via convolutional networks in both training and testing processes.
}


\subsection{Question Boosting}
\begin{figure}%
\centerline{\includegraphics[width=0.85\textwidth]{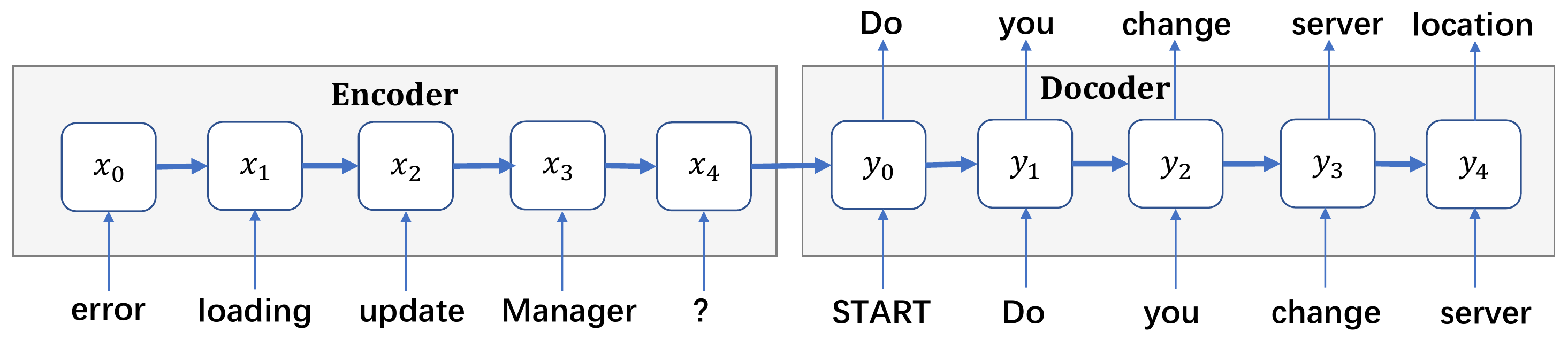}}
\caption{Question boosting process}
\label{fig:question_boosting}
\end{figure}

The task of question boosting is to automatically generate clarifying questions from the title of an initial post. This can be formulated as a sequence-to-sequence learning problem. Given $\mathbf{Q}$ is a sequence of words within the question title of an initial post, our target is to generate a useful clarifying question $\mathbf{CQ}$, which is relevant, syntactically and semantically correct. To be more specific, the goal is to train a model $\theta$ using $\langle$\textit{q}, \textit{cq}$\rangle$ pairs such that the probability $P_{\theta}(\mathbf{CQ}|\mathbf{Q})$ is maximized over the given training dataset. 
Mathematically, this query boosting task is defined as finding $\overline{y}$, such that:
\begin{equation}
    \overline{y} = argmax_{\mathbf{CQ}}P_{\theta}(\mathbf{CQ}|\mathbf{Q})
\end{equation}
$P_{\theta}(\mathbf{CQ}|\mathbf{Q})$ can be seen as the conditional log-likelihood of the clarification question $\mathbf{CQ}$ given the input post $\mathbf{Q}$. The encoder-decoder architecture has been used in addressing such a problem. We demonstrate an example of the question boosting process in Fig~\ref{fig:question_boosting}. 
The original post title ``error loading update manager ?'' is fed into the encoder, and the clarifying question ``do you change server location ?'' is the decoder target output. 

\subsubsection{Encoder}
The sequence of words within a post title is fed sequentially into the encoder, which generates a sequence of hidden states. Our encoder is a two-layer bidirectional LSTM network,
\begin{equation}
\nonumber
\begin{split}
\overrightarrow{\mathbf{f}{w}_{t}} &= \overrightarrow{{\textnormal{LSTM}_2}} \left( x_{t}, \overrightarrow{\mathbf{h}_{t-1}}\right) \\ 
\overleftarrow{\mathbf{b}{w}_{t}} &= \overleftarrow{{\textnormal{LSTM}_2}} \left( x_{t}, \overleftarrow{\mathbf{h}_{t-1}}\right) 
\end{split}
\end{equation}
where $x_{t}$ is the given input word token at time step $t$, and $\overrightarrow{\mathbf{h}_{t}}$ and $\overleftarrow{\mathbf{h}_{t}}$ are the hidden states at time step $t$ for the forward pass and backward pass respectively. The hidden states (from the forward and backward pass) of the last layer of the encoder are concatenated to form a state $s$ as  $\mathbf{s} = [\overrightarrow{\mathbf{f}{w}_{t}}; \overleftarrow{\mathbf{b}{w}_t} ]$.

\subsubsection{Decoder}
Decoder is singe layer LSTM network, initialized with the state $s$ as  $\mathbf{s} = [\overrightarrow{\mathbf{f}{w}_{t}}; \overleftarrow{\mathbf{b}{w}_t} ]$. Let $qword_{t}$ be the target word at time stamp $t$ of the clarifying question. During training, at each time step $t$ the decoder takes as input the embedding vector $y_{t-1}$ of the previous word $qword_{t-1}$ and the previous state $s_{t-1}$, and concatenates them to produce the input of the LSTM network. The output of the LSTM network is regarded as the decoder hidden state $s_t$, as follows:
\begin{equation}
\mathbf{s}_t = \textnormal{LSTM}_1 \left( y_{t-1} , \mathbf{s}_{t-1}\right) 
\end{equation}
The decoder produces one symbol at a time and stops when the \emph{END} symbol is emitted. The only change with the decoder at testing time is that it uses output from the previous word emitted by the decoder in place of $word_{t-1}$(since there is no access to a ground truth then).

\subsubsection{Attention Mechanism}
To effectively align the target words with the source words, we model the attention \cite{bahdanau2014neural} distribution over words in the target sequence. We calculate the attention $(a^{t}_{i})$ over the $i^{th}$ input token as :

\begin{equation}
    e^{t}_{i} = v^{t}\textnormal{tanh}\left(W_{eh}h_{i} + W_{sh}s_{t} + b_{att} \right)
\end{equation}
\begin{equation}
     a^{t}_{i} = \textnormal{softmax} \left( e^{t}_{i} \right) 
\end{equation}

Here $v^{t}$, $W_{sh}$ and $b_{att}$ are model parameters to be learned, and $h_{i}$ is the concatenation of forward and backward hidden states of source-code encoder. We use this attention $a^{t}_{i}$ to generate the context vector $c^{*}_{t}$ as the weighted sum of encoder hidden states : 

\begin{equation}
\mathbf{c}^{*}_{t} = \sum_{i=1,..,|\mathbf{x}|} a^{t}_{i} \mathbf{h}_i
\end{equation}
We further use the $c^{*}_{t}$ vector to obtain a probability distribution over the words in the vocabulary as follows, 

\begin{equation}
P = \textnormal{softmax} \left(\mathbf{W}_{v}[s_{t}, c^{*}_{t}] + b_{v} \right)
\end{equation}
where $W_{v}$ and $b_{v}$ are model parameters. Thus during decoding, the probability of a word is $P(qword)$. During the training process for each word at each timestamp, the loss associated with the generated question is : 

\begin{equation}
Loss = -\frac{1}{T} \sum^{T}_{t=0}logP(qword_{t})
\end{equation}

Once the model is trained, we do inference using beam search~\cite{graves2012sequence} and append the generated clarifying question to the original post title. The beam search is parameterized by the possible paths number $k$. The inference process stops when the model generates the END token, which stands for the end of the sentence. 

\begin{figure*}%
\centerline{\includegraphics[width=0.95\textwidth]{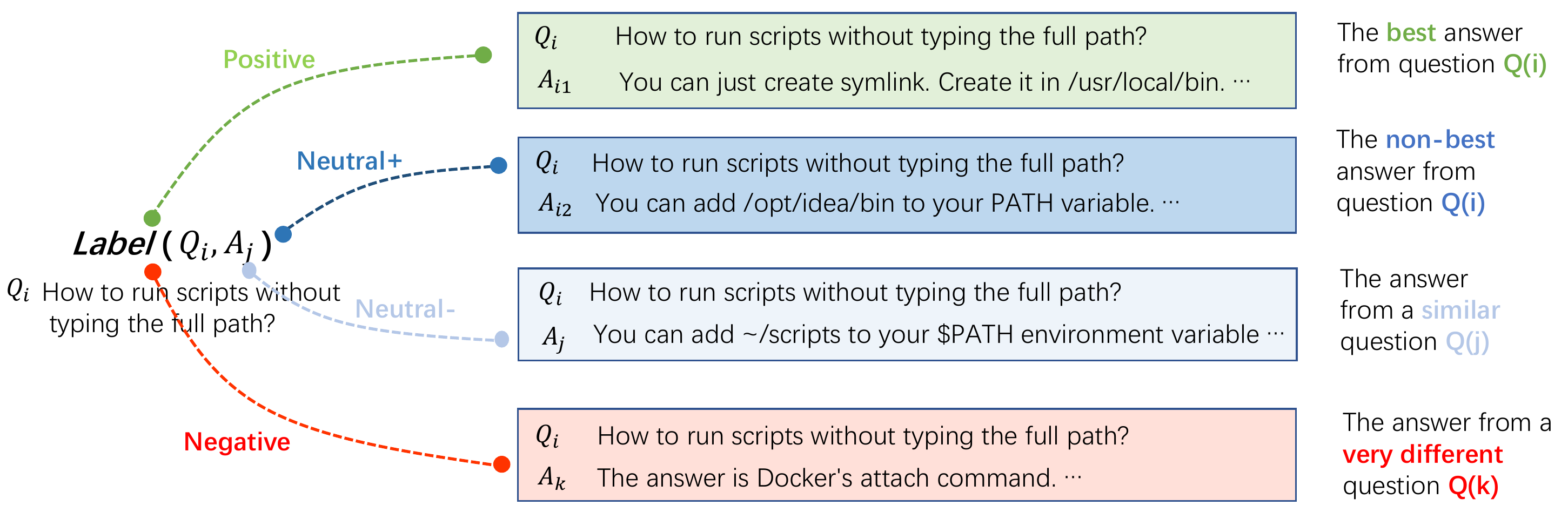}}
\caption{Label Establishing Process}
\label{fig:labeling}
\end{figure*}

\subsection{Label Establishment}
\label{subsec:label_establish}
According to our empirical study results from Section 2, the \revv{\emph{answer hungry}} phenomenon widely exists in technical Q\&A forums, i.e. only a small proportion of questions have an ``resolved'' 
answer, while many others remain unanswered and/or unresolved. Due to the reason of professionality of technical questions, only the experts with specific knowledge are qualified to evaluate the matching degree between a question and an answer. Therefore it is very hard to find such annotators and/or the creation of training sets requires a substantial manual effort. To address such a problem, We propose a novel scheme to automatically labeling each QA pair as \emph{positive}, \emph{neutral$^+$}, \emph{neutral$^-$}, and \emph{negative} samples. Fig~\ref{fig:labeling} shows an example of our labeling process. We propose four heuristic rules to label the QA pairs:


\begin{itemize}
    \item \emph{Positive samples}: for a given question $Q_i$, we pair it with its marked "best" answer (if it has one) $A_{i1}$, and label this qa pair as \emph{Positive}.
    \item \emph{Neutral$^+$ samples}: for a given question $Q_{i}$, we pair it with its non-best answer (answers within the same question thread, except the one marked as the best answer), and label this qa pair as \emph{Neutral$^+$}.
    \item \emph{Neutral$^-$ samples}: for a given question $Q_{i}$, we randomly select one answer $A_{j}$ from questions similar to it, 
    then label this question-answer pair as \emph{Neutral$^-$}. 
    \item \emph{Negative samples}: for a given question $Q_{i}$, we pair it with a randomly selected answer $A_{k}$ from non-similar questions and label this QA pair as \emph{Negative}.
\end{itemize}
\rev{Since we are recommending answers from candidate answers of questions relevant to the query question, if the retrieved questions are not relevant to the query question, it is unlikely we can select the best answer from the answer candidates pool. We followed the question retrieval method proposed by Xu et al.~\cite{xu2017answerbot} to search for similar questions, which has been proven to be more effective for this task of relevant question retrieval. We used the IDF-weighted word embedding to calculate the similarity score between the query and the question title. Thereafter, a set of similar questions can be identified by selecting the top-k ranked questions.} 

After this label establishing process, we can gather large amounts of labeled examples, which greatly saves the time-consuming and labor-intensive labeling process. 

\subsection{Answer Recommendation}
After collecting large amounts of labeled training data via label establishment, we are able to train the deep learning model based on the four kinds of training samples.

\begin{figure*}%
\centerline{\includegraphics[width=0.95\textwidth]{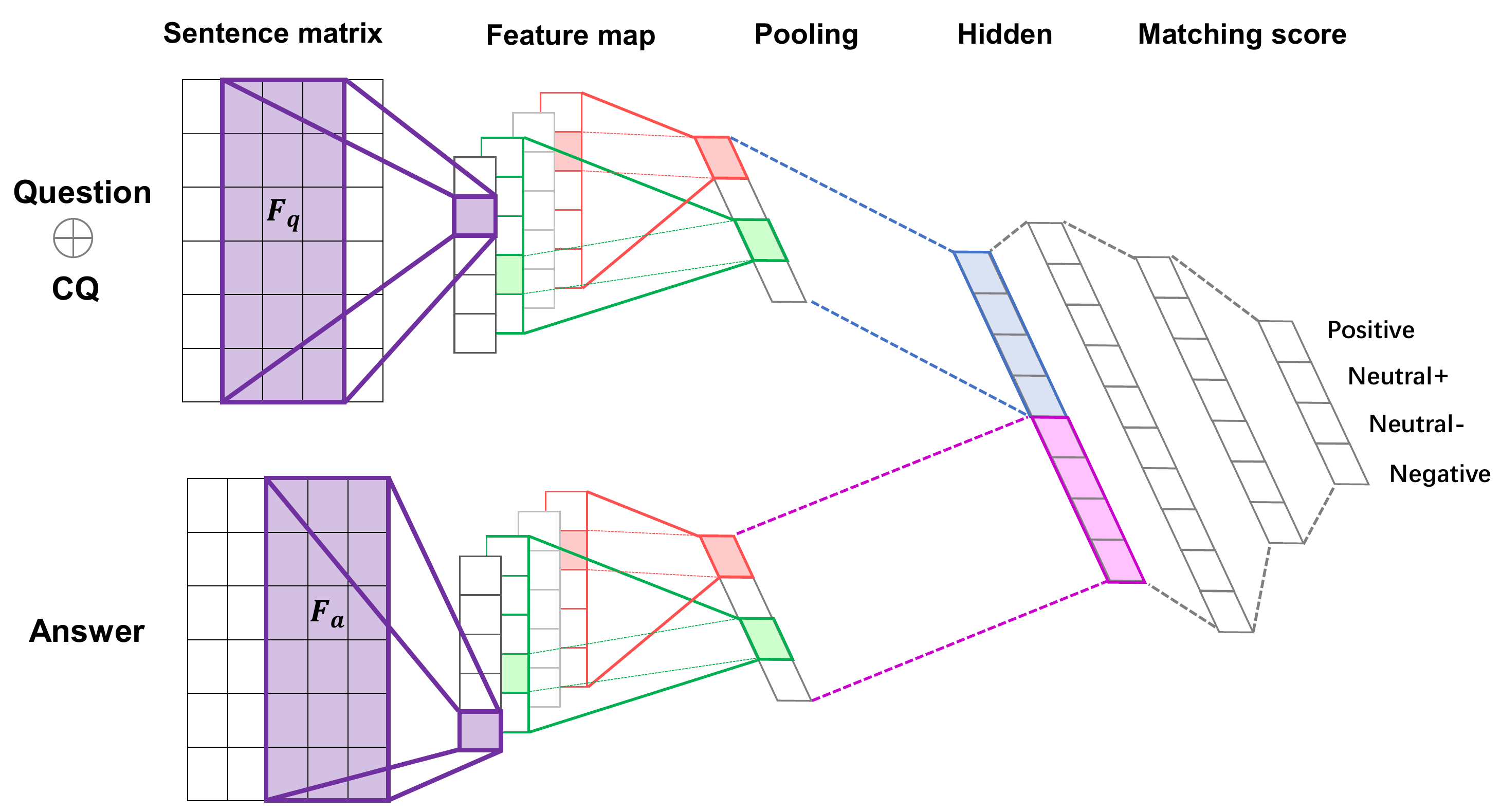}}
\caption{Overall architecture of the answer recommendation model.}
\label{fig:workflow}
\end{figure*}

We present a weakly supervised neural network architecture for ranking QA pairs. Fig.~\ref{fig:workflow} demonstrates the workflow of our proposed model.
The main building blocks of our architecture are two convolutional neural networks~\cite{kim2014convolutional, kalchbrenner2014convolutional}. These two underlying sub-models work in parallel, mapping questions and answers to their distributional vectors respectively, which are then used to calculate the final similarity score between them. 

\subsubsection{Sentence Matrix} \rev{The input to our model are $\langle q \oplus cq, a\rangle$ pairs, where $q$ and $a$ stands for the question and answer of a labelled QA pair, $cq$ stands for the clarifying questions generated by our question boosting model. The questions (including the original questions and clarifying questions) and answers are parallel sentences,}
where each sentence is treated as a sequence of words: $(w_1,...,w_s)$, where each word is drawn from a vocabulary $\mathbf{V}$. Words are represented by distributional vectors 
$\mathbf{w} \in \mathbb{R}^{1 \times d}$ via looking up in a pre-trained word embedding matrix $\mathbf{W} \in \mathbb{R}^{d \times |V|}$. 

For each input $\langle q \oplus cq, a\rangle$ pair, we build two sentences matrix $\mathbf{S_q}$ and $\mathbf{S_a} \in \mathbb{R}^{d \times |s|}$ for each question and answer respectively, where the $i$th column represents the word embedding of $\mathbf{w_i}$ at corresponding position $i$ in a sentence. 
\subsubsection{Convolutional feature maps}
To learn to capture and compose features of individual words in a given sentence from low-level word embeddings into higher level semantic concepts, we apply two identical convolutional neural network blocks to the input sentence matrix $\mathbf{S_q}$ and $\mathbf{S_a}$ respectively. 


More formally, the convolution operation $*$ between an input sentence matrix $\mathbf{S_{q/a}} \in \mathbb{R}^{d \times |s|} $ and a filter $\mathbf{F} \in \mathbb{R}^{d \times m}$ (called a filter of size $m$) results in a vector $\mathbf{c} \in  \mathbb{R}^{|s|-m+1}$, where each component is computed as follows:
\begin{equation}
    \mathbf{c_{i}} = \left(\mathbf{S} * \mathbf{F} \right)_{i} = \sum_{k, j} \left( \mathbf{ S_{[:,i-m+1:i]} } \otimes \mathbf{F} \right)_{kj}
\end{equation}

In the above equation,  $\otimes$ is the element-wise multiplication and $\mathbf{S_{[:,i-m+1:i]}}$ is a matrix slice of size $m$ along the columns. Note that the convolution filter is of the same  dimensionality $d$ as the input sentence matrix. As shown in Fig.~\ref{fig:workflow}, it slides along the column dimension of $\mathbf{S}$ producing a vector $\mathbf{c} \in  \mathbb{R}^{|s|-m+1}$. Each component $\mathbf{c_{i}}$ is the result of computing an element-wise product between a column slice of $\mathbf{S}$ and the filter matrix $\mathbf{F}$, which is then flattened and summed producing a single value. By applying a set of filters (called a filter bank) $\mathbf{F} \in \mathbb{R}^{n \times d \times m}$ to sequentially convolved with the sentence matrix $\mathbf{S}$ will generate a convolutional feature map matrix $\mathbf{C \in \mathbb{R}^{n \times (|s|-m+1)}}$. 

\subsubsection{Pooling layer} 
Following that, we pass the output from the convolutional layer to the pooling layer, whose goal is to aggregate the information and reduce the representation. We apply a max pooling operation \cite{collobert2011natural} over the convolutional feature map and take the maximum value $\mathbf{\widehat{c}} = max\{ \mathbf{c_{i}} \}$ as the feature corresponding to a particular filter. The idea is to capture the most important feature - one with the highest value - for each feature map.


\subsubsection{Matching score layer}
The output of the penultimate convolutional and pooling layers $x$ is passed to a series of fully connected layer followed by a softmax layer. It computes the probability distribution over the four kinds of labels (\emph{positive}, \emph{neutral$^+$}, \emph{neutral$^-$}, \emph{negative}):
\begin{equation}
    P\left(y=j| \mathbf{x} \right) = \frac{ e^{\mathbf{x}^T\theta_j} }{ \sum_{k=1}^{K} e^{\mathbf{x}^T\theta_k} }
\end{equation}
where $\theta_{k}$ is a weight vector of the $k$-th class. $\mathbf{x}$ can be thought of as a final abstract representation of the input QA pair obtained by a series of transformations from the input layer through a series of convolutional and pooling operations.

For the final matching score, we want this score to be high if the input qa pair is \emph{positive} and \emph{neutral$^+$}, and to be low if it is \emph{negative} and \emph{neutral$^-$}. Therefore we define the calculation of  the   similarity score as follows:
\begin{equation}
\label{eq:score_eq}
    Score = \omega_{pos} \times P\left(\emph{pos} \right) 
    + \omega_{neu+} \times P\left(\emph{neu$^+$} \right) 
    - \omega_{neu-} \times P\left(\emph{neu$^-$} \right) 
    - \omega_{neg}  \times P\left(\emph{neg} \right)
\end{equation}

\begin{algorithm}[H]
\label{Alg:offline}
\SetKwInput{KwOffInput}{Input}             
\SetKwInput{KwOffOutput}{Output}      
\SetAlgoLined
 \KwOffInput{Data dump of technical Q\&A sites\;}
 \KwOffOutput{1.Question Boosting model; 2.Answer recommendation model\;}
 Extract $\langle q, cq \rangle$ pairs from data dump\;
 Train $\langle q, cq \rangle$ pairs with attentional-based seq2seq model \;
 Save the model as Question Boosting model \;
 Extract $\langle q, a \rangle$ pairs from data dump\;
 \For{$q_{i}, a_{i} \in \langle q, a \rangle pairs$}
 {
    \If{$q_{i}$ has accepted-answer} {
        \If{$a_{i}$ is accepted-answer} {
            Label $\langle q_{i}, a_{i} \rangle$ as $Positive$ \;
        }
        \Else {
            Label $\langle q_{i}, a_{i} \rangle$ as $Neutral^{+}$ \;
        }
        Select similar answer $a_{j}$ then Label $\langle q_{i}, a_{j} \rangle$ as $Neutral^{-}$ \;
        Select random answer $a_{k}$ then Label $\langle q_{i}, a_{k} \rangle$ as $Negative$ \;
    }
 }
 \For{$q_{i} \in labelled \langle q, a \rangle  pairs$}{
    Generate $cq_{i}$ for $q_{i}$ using Question Boosting model\;
    Append $cq_{i}$ to $q_{i}$ to make labelled $\langle q_{i} \oplus cq_{i}, a_{i} \rangle$ pair \;
 }
 Train labelled $\langle q \oplus cq, a \rangle$ pairs with CNN-based classification model \;
 Save the model as Answer Recommendation model
 \caption{\minor{DeepAns Algorithm (Offline Training)}}
\end{algorithm}
\vspace{-0.cm}

\begin{algorithm}[H]
\label{Alg:online}
\SetKwInput{KwOnInput}{Input}             
\SetKwInput{KwOnOutput}{Output}      
\SetAlgoLined
 \KwOnInput{User search query $q_{user}$\;}
 \KwOnOutput{A set of candidate answers with a matching score for each answer \;}
 Generate $cq$ for $q_{user}$ using Question Boosting model \;
 Search top-k similar questions for the given query $q_{user}$ \;
 Add top-k questions to similar question set $SQ$ \;
 \For{$q_{i} \in SQ$} {
    \For{$a_{j} \in q_{i}$} {
        Add answer to candidate answers set $CA$ \;
    }
 }
 \For{$a_{i} \in CA$} {
    Pair $a_{i}$ with expanded query to make a $\langle q_{user} \oplus cq, a_{i} \rangle$ pair \;
    Fit $\langle q_{user} \oplus cq, a_{i} \rangle$ pair to Answer Recommendation model \;
    Compute the final matching score $s_{i}$ via Equation~\ref{eq:score_eq}
 }
 \caption{\revv{DeepAns Algorithm (Online Recommendation)}}
 Rerank answers in $CA$ via matching scores
\end{algorithm}

There are four weights as shown in Equation~\ref{eq:score_eq}. We initially set all the four weights to 1 at the beginning. Then the optimal settings of these weights are carefully tuned on our validation set (detailed in Section~\ref{subsec:ensitivity}). We use the final matching score to measure the relevance between a question and an answer.

\subsection{\revv{DeepAns Algorithm}}
\revv{
We divide our model into two components: offline training and online recommendation. 
The detailed algorithms of \emph{DeepAns} for offline training and online recommendation are presented in Algorithm~\ref{Alg:offline} and Algorithm~\ref{Alg:online} respectively. 
To be more specific, during the offline training, we use the data from technical Q\&A sites to train the question boosting model (lines 1-3) and answer recommendation model (lines 4-20). 
When it comes to the online recommendation, for a given user query, we first collect a pool of answer candidates via finding its similar questions (lines 1-8).  
After that, we use the trained question boosting model to perform query expansion, then pair it with each of the answer candidates and fit them into the trained answer recommendation model to estimate their matching scores (lines 9-14). 
}

%% file: eval.tex
In this section, we first describe the data sets used throughout our experiments. We then discuss the baselines we compare to our new \emph{DeepAns} approach and our experimental settings. Lastly, we explain the automatic evaluation process.



\subsection{Data Preparation}
We collected data from the official dump of StackExchange, a network of online question and answer websites.
The StackExchange data dump contains timestamped information about the posts, comments as well as the revision history made to the post. Each post comprises a short question title, a detailed question body, corresponding answers and multiple tags. For each post, users can add clarifying questions to posts for further discussion. After receiving one or more answers, the asker can select one answer that is most suitable for their question as the accepted/best answer. 
\revv{We choose three different technical Q\&A sites, i.e., Ask Ubuntu, Super User and Stack Overflow for our experiment. These three technical Q\&A sites are commonly used by software developers and each one focuses on a specific area. For instance, Ask Ubuntu and Super User focus on Ubuntu system questions and computer software/hardware questions respectively, and Stack Overflow is the most popular programming related Q\&A site which has been heavily used by software developers via the internet. As with our previous empirical study, we only focus on the Python and Java related questions in Stack Overflow for this study, referred to as SO (Python) and SO (Java) respectively in this study. }


\begin{table*} \vspace{-0.0cm}
\caption{\revv{Number of Training/Validation/Testing Samples}}
\revv{
\begin{center}
\vspace{-0.2cm}\begin{tabular}{||c|l|c|l|c||}
    \hline
    \multirow{3}{*}{Ask Ubuntu}  
                              & \# $\langle q, cq \rangle$ pairs & 68,216 & \# $\langle q, a \rangle$ pairs & 289,062 \\ \cline{2-5}
                              & \# \emph{Positive} pairs & 79,726 & \# \emph{Neutral$^+$} pairs & 49,884 \\ \cline{2-5}
                              & \# \emph{Neutral$^-$} pairs & 79,726  & \# \emph{Negative} pairs & 79,726  \\ \cline{2-5}
    \hline
    \multirow{3}{*}{Super User}  
                              & \# $\langle q, cq \rangle$ pairs & 87,081  & \# $\langle q, a \rangle$ pairs & 447,221 \\ \cline{2-5}
                              & \# \emph{Positive} pairs & 119,305 & \# \emph{Neutral$^+$} pairs & 89,306 \\ \cline{2-5}
                              & \# \emph{Neutral$^-$} pairs & 119,305 & \# \emph{Negative} pairs & 119,305 \\ \cline{2-5}
    \hline\hline
    \multirow{3}{*}{SO (Python)}  
                              & \# $\langle q, cq \rangle$ pairs & 311,127 & \# $\langle q, a \rangle$ pairs & 2,372,232 \\ \cline{2-5}
                              & \# \emph{Positive} pairs & 610,948 & \# \emph{Neutral$^+$} pairs & 539,388 \\ \cline{2-5}
                              & \# \emph{Neutral$^-$} pairs & 610,948 & \# \emph{Negative} pairs & 610,948 \\ \cline{2-5}
    \hline
    \multirow{3}{*}{SO (Java)}  
                              & \# $\langle q, cq \rangle$ pairs & 456,077 & \# $\langle q, a \rangle$ pairs & 3,013,859 \\ \cline{2-5}
                              & \# \emph{Positive} pairs & 734,977 & \# \emph{Neutral$^+$} pairs & 808,928 \\ \cline{2-5}
                              & \# \emph{Neutral$^-$} pairs & 734,977 & \# \emph{Negative} pairs & 734,977 \\ \cline{2-5}
    \hline
\end{tabular}
\label{tab:dataoverview}
\end{center}
}
\vspace{-0.0cm}
\end{table*}

\rev{The experimental dataset creation process is divided into three phases: extracting $\langle q, cq \rangle$ pairs, constructing labelled $\langle q, a \rangle$ pairs, and constructing labelled $\langle q \oplus cq, a \rangle$ pairs}, where $q$ stands for the question, $cq$ stands for the clarifying question, and $a$ stands for the answer. Table~\ref{tab:dataoverview} describes the statistics of our collected datasets.
\begin{enumerate}
    \item \noindent \textbf{Extract $\langle q, cq \rangle$ Pairs}: For each post, 
    \rev{we follow the methods described in Section~\ref{subsec:motivation_cqs} to extract the clarifying questions. According to our manual analysis results, we summarize a list of keywords associated with non-clarifying questions, such as ``edit'', ``related'', ``vote'', etc. We preprocess our dataset to remove all instances that involve such keywords. We also summarize a list of key phrases associated with the clarifying questions, such as ``do you'', ``have you'', ``how'', ``which'', etc. We retained the pairs that include the above key phrases. After that,} we pair the original post with its associated clarifying question as $\langle q, cq \rangle$ pairs. We extract a total of 68,216 pairs in Ask Ubuntu, and 87,081 pairs in Super User. \revv{The number of $\langle q, cq \rangle$ pairs in Stack Overflow are much larger, we obtain a total of 311,127 pairs for SO (Python) and  456,077 pairs for SO (Java).}
    These collected $\langle q, cq \rangle$ pairs are used to train a sequence-to-sequence model for question boosting.
    \item \noindent \textbf{Construct labelled $\langle q, a \rangle$ Pairs}: To make the $\langle q, a \rangle$ pairs, we first extract the questions that having explicitly marked accepted answers. Then for each question, we pair it with the accept answer to make the \emph{positive sample}, with non-accepted answer to make the \emph{neutral$^+$ sample}, with an answer to a similar question to make the \emph{neutral$^-$ sample}, and with an answer to a randomly selected question to make the \emph{negative sample}. 
    \revv{We have to clarify that some questions do not have the non-accepted answers; this is the reason why the number of \emph{neutral$^+$} samples is smaller than the number of other samples,
    such as Ask Ubuntu, Super User and SO (Python), while some other questions have more than one non-accepted answers, which results in the number of \emph{neutral$^+$} samples is bigger than those of the rest, such as SO (Java). 
    For the final dataset, we construct 289,062 and  447,221 $\langle q, a \rangle$ labelled pairs for Ask Ubuntu and Super User, and 2,372,232 and 3,013,859 $\langle q, a \rangle$ labelled pairs for SO (Python) and SO (Java) respectively.
    It is obvious that the number of qa pairs in Stack Overflow far outnumber those of other technical Q\&A sites.  
    After the label establishment process, we largely expand the labelled dataset for training. We randomly sample 5,000 questions for validation and 5,000 questions for testing respectively, and kept the rest for training.
    It is worth mentioning that we first used the validation set for model selection regarding the accuracy of QA pairs classification results, which is a middle result of the answer selection target. After that, we reused the validation set for tuning the four weights as shown in Equation~\ref{eq:score_eq}. The testing set was used only for testing the final solution to confirm the actual predictive power of our model with optimal parameter settings.
    } 
    \item \noindent \textbf{Construct labelled $\langle q \oplus cq, a \rangle$ Pairs}: \rev{For each labelled $\langle q, a \rangle$ pair, we feed the original question to the trained question boosting model to generate a clarifying question.
    After that, we append the clarifying question to the original question to construct the $\langle q \oplus cq, a \rangle$ pairs. 
    The number of the $\langle q \oplus cq, a \rangle$ pairs is identical with the number of  $\langle q, a \rangle$ pairs.}
\end{enumerate}


    

\subsection{Implementation Details}
We implemented our \emph{DeepAns} system in Python using the PyTorch framework. The main parameters of our deep learning model (tuned using the validation dataset) were as follows: 
\begin{itemize}
    \item Question Boosting: 
    \revv{
    We train an attentional sequence-to-sequence model for this subtask.
    Previous studies have shown that the deep sequence-to-sequence model can achieve state-of-the-art performance on different tasks~\cite{sennrich2016edinburgh, iyer2016summarizing, hu2018deep, gao2020generating}.
    We also used the parameter settings from~\cite{gao2020generating} for training the $\langle q, cq \rangle$ pairs in this study. We use a two-layer bidirectional LSTM for the encoder and a single-layer LSTM for the decoder. We set the number of LSTM hidden states to be 256 in both encoder and decoder. Optimization is performed using stochastic gradient descent (SGD) with a learning rate of 0.01. During decoding, we perform beam search with a beam size of 10.
    }
    \item Answer Recommendation: 
    \revv{
    Kim et al.~\cite{kim2014convolutional} have shown that convolutional neural networks trained on top of pre-trained word vectors achieved promising performance for sentence-level classification tasks.
    Hence in our work, we also followed the experiment settings of their studies. 
    We initialize the word embeddings from our unsupervised corpus and set the dimension of word embedding $d$ to 100. The width $m$ of the three convolution filters is set to 3, 4, 5 and the number of convolution feature maps is set to 100. We use ReLu activation function and a simple max-pooling function. The size of the hidden layer is equal to the size of the join vector obtained after concatenating question and answer vectors from the distributional models.
    }
\end{itemize}

To train both networks, we used stochastic gradient descent with shuffled mini-batches. The batch size is set to 64. Both network are trained for 50 epochs with early stopping, i.e., we stop the training if no update to the best accuracy on the validation set has been made for the last 5 epochs.

\subsection{Baselines}
To demonstrate the effectiveness of our proposed {\sc DeepAns}, we compared it with several comparable systems. We briefly introduced these and how they are used for the task of predicting the best answer among a set of answer candidates. 
\rev{{\sc DeepAns} is built with the semantic features of words in their dimensions, we used the average word vector of a sentence as features for training all of the baseline models for a fairer comparison.}
For each baseline method, their parameters were carefully tuned, and the parameters with the best performance were used to report the final comparison results with our \emph{DeepAns} approach on the same datasets:

\begin{itemize}
    \item \textbf{Learning to Rank} The answer prediction problem of our task is similar to the traditional ranking task \cite{savenkov2015ranking} \cite{agarwal2012learning}, where the given question and a set of answer candidates are analogous to a query and a set of relevant entities. Hence our task is transformed to find an optimal ranking order of these answer candidates according to their relevance to a given question. We choose the AdaRank\cite{xu2007adarank} and LambdaMART \cite{burges2010ranknet} as the baseline learning-to-rank methods for our task. 
    \rev{We used the \emph{positive}, \emph{neutral$+$} as the target value to define the order of each example. This is reasonable because the label establishment is part of our model, and the heuristic rules for setting up the \emph{neutral$-$} and \emph{negative} samples are never used before.
    }

    \item \textbf{Traditional Classifiers} Recently Calefato et al.~\cite{calefato2019empirical} proposed to approach the best answer prediction problem as a binary-classification task, and in their work they assessed 26 best-answer prediction classifiers in Stack Overflow. We choose the two most effective traditional classifiers from their experimental results, xgbTree and RandomForest, for use in our study. \rev{
    As they were doing binary classification, to adapt to our training data, we kept our \emph{positive samples} as positive and consider 
    \emph{neutral$^+$} samples as negative.
    Thereafter, we utilize the classification models to generate an answer ranking list by pairwise comparison between the answer candidates.}

    \item \textbf{AnswerBot} 
    Xu et al.~\cite{xu2017answerbot} proposed a framework called AnswerBot to generate an answer summary for a non-factoid technical question. Their user study showed a promising performance for selecting salient answers by their method. \revv{We adapted their AnswerBot approach for our task of recommending answers among a set of answer candidates. To be more specific, for a given question, AnswerBot generates a ranked list of candidate answers according to the ranking scores. This ranked list of answers is then used to calculate the precision of answer selection results.
    }

    \revv{
    \item \textbf{IR-DeepAns} To verify the effectiveness of using clarifying questions as a way of question boosting, compared with our sequence to sequence model, we also considered a simple IR-based approach using similar clarifying questions as a query expansion mechanism. 
    For a given question $q_{i}$, we first identified the most similar question $q_{j}$ in $\langle q, cq \rangle$ dataset, and then retrieved the clarifying question $cq_{j}$ associated with $q_{j}$. We applied IDF-weighted word embedding methods to calculate the similarity score between two questions. We feed the $q_{i}$ and $cq_{j}$ into our model and name this baseline as IR-DeepAns. This model is close to ours. 
    }

\end{itemize}

\subsection{Evaluation Methods}
\subsubsection{\revv{Experiment Setup}}
\revv{
To thoroughly evaluate our model, we conducted a large-scale automatic evaluation experiment. We used IDF-weighted word embedding (described in Section~\ref{subsec:label_establish}) to calculate the similarity score between two question titles.
For each testing qa pair $\langle q_{t}, a_{t} \rangle$, we then performed K-NN (K=5) to search for similar questions over the whole data set for the given testing question $q_{t}$. We then constructed an answer candidate pool by gathering the top-5 answers associated with these selected questions. Since the top-similar question extracted by K-NN is always the original post itself, we can ensure that the accepted answer $a_{t}$ paired with the original post $q_{t}$ is always in the answer candidate pool. In other words, the answer candidate pool for testing question $q_{t}$ contains 5 answers, one of which is the accepted answer $a_{t}$.
In summary, for the 5,000 testing questions of each platform, we constructed 5,000 $\times$ 5 QA pairs in total to serve as the final evaluation sets. 
Following this, for each testing question $q_{t}$, we first applied the pre-trained question boosting model to generate a clarifying question $cq_{t}$. We then paired the given question with each answer in the candidate pool to construct the $\langle q_{t} \oplus cq_{t}, a_{t} \rangle$ pairs. 
The $\langle q_{t} \oplus cq_{t}, a_{t} \rangle$ pair was fitted into our model to calculate a matching score, and we then generated a ranking order for each group of candidate answers according to their matching scores to the given question.
} 

\subsubsection{\revv{Evaluation Metrics}}
Since the evaluation answer candidate pool includes the accepted answer, one way to evaluate our approach is to look at how often the accepted answer is ranked higher up among members of the answer candidate pool.
Thus we adopted the widely-accepted metric, $P@K$ and $DCG@K$ to measure the ranking performance of our model. 
\begin{itemize}
    \item $P@K$ is the precision of the best answer in top-K candidate answers. \rev{Given a question, if one of the top-k ranked answers is the best answer, we consider the recommendation to be successful and set $success(best_i \in topK)$ to 1, otherwise, we consider the recommendation to be unsuccessful and set $success(best_i \in topK)$ to 0. The $P@K$ metric is defined as follows:} 
    \begin{equation}
    P@K = \frac{1}{N}\sum_{i=1}^N[success(best_i \in topK)]
    \end{equation}
    \item $DCG@K$ is \rev{another popular top-K accuracy metric that measures a recommender system performance based on the graded relevance of the recommended items and their positions in the candidate set. Different from $P@K$, the intuition of $DCG@K$ is that highly-ranked items are more important than low-ranked items. According to this metric, a recommender system gets a higher reward for ranking the correct answer at a higher position. 
    The $success(best_i \in topK)$ is same with the previous definition, while the $rank_{best_i}$ is the ranking position of the best answer $i$. The $DCG@K$ is defined as follows:}
    \begin{equation}
    DCG@K = \frac{1}{N}\sum_{i=1}^N\frac{[ success(best_i \in topK) ]}{\log_2(1+rank_{best_i})}
    \end{equation}
\end{itemize}

%% file: results.tex
To gain a deeper understanding of the performance of our approach, we conducted an analysis on our large-scale automatic evaluation results. 
Specifically, we mainly focus on the following research questions:
\begin{itemize}
    \item \textit{RQ-1:} How effective is our approach under automatic evaluation?
    \item \textit{RQ-2:} How effective is our use of  \emph{Question boosting} and \emph{Label establishing} methods?
    \item \textit{RQ-3:} \revv{How effective is our approach under different parameter settings?} 
\end{itemize}

\subsection{\textbf{RQ-1: Automatic Evaluation Results Analysis}}
The automatic evaluation results of our proposed model and aforementioned baselines over \revv{different technical Q\&A sites are summarized in Table~\ref{tab:automatic_eval_ask}, Table~\ref{tab:automatic_eval_super}, Table~\ref{tab:automatic_eval_python} and Table~\ref{tab:automatic_eval_java} respectively.} 
We do not report $P@5$ and $DCG@1$ in our tables, since $DCG@1$ is always equal to $P@1$ and $P@5$ will always be equal to 1. 
The best performing system for each column is highlighted in boldface. As can be seen, \textbf{our model outperforms all the other methods by a large margin} in terms of $P@K$ score and $DCG@K$ score. From the table, we can observe the following points discussed below.

\begin{enumerate}
    \item Compared to traditional classifiers, such as xbgTree and RandomForest, one can clearly see that our approach performs much better. For example, it improves over xgbTree on $P@1$ by 42\% on Ask Ubuntu dataset, and 39\% on Super User dataset. 
    \item Compared with the method proposed by \cite{calefato2019empirical}, which only has two kinds of labels (\emph{positive} and \emph{negative}), our approach constructs four kinds of labeled data (\emph{positive}, \emph{neutral$^+$}, \emph{neutral$^-$}, \emph{negative}) automatically via incorporating the \emph{label establishing} process. By introducing the \emph{neutral$^+$} and \emph{neutral$^-$} training samples, our approach can learn how to separate the best answer from the similar ones, which may explain the obvious advantage of our model in $P@1$. 
    \item Our approach also outperforms the AnswerBot by a large margin. We attribute this to the following reasons. Firstly, by adding a clarifying question into our model, we can properly fuse the information between the isolated question sentences and answers, which can reduce the lexical gap between them and better pair the answer with associated questions. Secondly, we use two parallel convolutional neural network block to learn optimal vector representation of QA pairs that preserving important syntactic and semantic features. To compute the matching score, we relate the rich representation features via a weakly supervised way from the available training data. 
    
\begin{table*}[htb]
\centering
\caption{Automatic evaluation (Ask Ubuntu)}
\label{tab:automatic_eval_ask}
\resizebox{0.95\textwidth}{!}{
    \begin{tabular}{||l|cccc|cccc||} 
      \hline
      Model & P@1  & P@2 & P@3 & P@4 & DCG@2 & DCG@3 & DCG@4 & DCG@5 \\
      \hline
      RandomForest & $26.6\pm1.6\%$ 
                  & $49.2\pm1.6\%$ 
                  & $70.8\pm1.6\%$ 
                  & $87.1\pm0.4\%$ 
                  & $40.9\pm1.5\%$ 
                  & $51.7\pm1.5\%$ 
                  & $58.8\pm0.9\%$ 
                  & $63.7\pm0.8\%$ \\
      XgbTree & $28.8\pm1.4\%$ 
            & $53.6\pm1.3\%$ 
            & $73.0\pm1.0\%$ 
            & $87.9\pm1.2\%$ 
            & $44.5\pm1.2\%$ 
            & $54.2\pm0.9\%$ 
            & $60.7\pm0.8\%$
            & $65.3\pm0.7\%$ \\ 
      LambdaMART & $25.4\pm1.1\%$ 
                  & $45.7\pm1.0\%$ 
                  & $65.7\pm1.2\%$ 
                  & $84.0\pm1.0\%$ 
                  & $38.5\pm1.0\%$ 
                  & $47.5\pm1.1\%$ 
                  & $55.8\pm0.9\%$ 
                  & $62.3\pm0.6\%$ \\
      AdaRank & $24.9\pm1.1\%$ 
            & $45.3\pm1.1\%$ 
            & $65.0\pm1.0\%$ 
            & $82.9\pm0.8\%$ 
            & $38.1\pm1.2\%$ 
            & $47.2\pm1.1\%$ 
            & $55.2\pm1.0\%$
            & $61.8\pm0.7\%$ \\ 
      AnswerBot  & $27.7\pm1.6\%$ 
               & $52.1\pm1.5\%$ 
               & $73.5\pm1.0\%$ 
               & $89.2\pm0.7\%$ 
               & $43.1\pm1.5\%$ 
               & $53.8\pm1.1\%$ 
               & $60.5\pm0.8\%$ 
               & $64.7\pm0.8\%$ \\
      \revv{DeepAns-IR}  & $37.2\pm2.0\%$ 
               & $59.9\pm2.1\%$ 
               & $77.5\pm1.7\%$ 
               & $92.0\pm1.0\%$ 
               & $50.8\pm1.5\%$ 
               & $59.6\pm1.3\%$ 
               & $65.8\pm1.1\%$ 
               & $68.7\pm0.8\%$ \\
      \hline
      \textbf{DeepAns}  & $\mathbf{40.9\pm1.5\%}$ 
                     & $\mathbf{61.7\pm1.9\%}$ 
                     & $\mathbf{77.9\pm0.9\%}$ 
                     & $\mathbf{92.0\pm0.9\%}$ 
                     & $\mathbf{54.0\pm1.7\%}$ 
                     & $\mathbf{62.1\pm1.1\%}$
                     & $\mathbf{68.2\pm1.1\%}$
                     & $\mathbf{71.3\pm0.9\%}$ \\ 
      \hline
\end{tabular}
}
\end{table*}

\begin{table*}[htb]
\centering
\caption{Automatic evaluation (Super-User)}
\label{tab:automatic_eval_super}
\resizebox{0.95\textwidth}{!}{
    \begin{tabular}{||l|cccc|cccc||} 
      \hline
      Model & P@1  & P@2 & P@3 & P@4 & DCG@2 & DCG@3 & DCG@4 & DCG@5 \\
      \hline
      RandomForest & $27.4\pm1.6\%$ 
                      & $50.2\pm1.7\%$ 
                      & $70.5\pm1.4\%$ 
                      & $87.3\pm0.9\%$ 
                      & $41.7\pm1.6\%$ 
                      & $51.9\pm1.4\%$ 
                      & $59.2\pm1.2\%$ 
                      & $64.1\pm0.9\%$ \\
      XgbTree & $29.2\pm1.6\%$ 
            & $55.9\pm1.3\%$ 
            & $74.2\pm0.9\%$ 
            & $88.5\pm0.7\%$ 
            & $47.6\pm1.2\%$ 
            & $56.7\pm1.1\%$ 
            & $63.0\pm0.9\%$
            & $67.4\pm0.8\%$ \\ 
      LambdaMART & $25.9\pm1.1\%$ 
                  & $47.1\pm1.0\%$ 
                  & $66.1\pm1.0\%$ 
                  & $84.5\pm1.2\%$ 
                  & $39.8\pm1.0\%$ 
                  & $48.8\pm1.0\%$ 
                  & $56.4\pm0.7\%$ 
                  & $62.9\pm0.6\%$ \\
      AdaRank & $25.1\pm1.2\%$ 
            & $46.2\pm1.1\%$ 
            & $65.7\pm1.0\%$ 
            & $84.1\pm0.9\%$ 
            & $38.4\pm1.1\%$ 
            & $47.6\pm1.1\%$ 
            & $55.7\pm1.0\%$
            & $62.2\pm0.9\%$ \\ 
      AnswerBot  & $29.8\pm1.4\%$ 
               & $53.9\pm1.2\%$ 
               & $74.1\pm1.3\%$ 
               & $89.6\pm0.8\%$ 
               & $45.0\pm1.2\%$ 
               & $55.1\pm0.9\%$ 
               & $61.8\pm0.7\%$ 
               & $65.8\pm0.6\%$ \\
       \revv{DeepAns-IR}  & $38.8\pm2.1\%$ 
               & $63.4\pm1.8\%$ 
               & $80.7\pm1.2\%$ 
               & $92.5\pm1.2\%$ 
               & $54.3\pm1.8\%$ 
               & $63.0\pm1.3\%$ 
               & $68.1\pm1.2\%$ 
               & $70.9\pm1.0\%$ \\
      \hline
      \textbf{DeepAns}  & $\mathbf{40.7\pm1.9\%}$ 
                     & $\mathbf{65.8\pm1.1\%}$ 
                     & $\mathbf{82.2\pm1.1\%}$ 
                     & $\mathbf{93.9\pm0.8\%}$ 
                     & $\mathbf{56.5\pm1.2\%}$ 
                     & $\mathbf{64.7\pm1.2\%}$
                     & $\mathbf{69.8\pm1.0\%}$
                     & $\mathbf{72.1\pm0.8\%}$ \\ 
      \hline
\end{tabular}
}
\end{table*}

\begin{table*}[htb]
\centering
\caption{\revv{Automatic evaluation (SO-Python)}}
\label{tab:automatic_eval_python}
\revv{
\resizebox{0.95\textwidth}{!}{
    \begin{tabular}{||l|cccc|cccc||} 
      \hline
      Model & P@1  & P@2 & P@3 & P@4 & DCG@2 & DCG@3 & DCG@4 & DCG@5 \\
      \hline
      RandomForest & $34.0\pm1.3\%$ 
                      & $57.2\pm1.0\%$ 
                      & $74.8\pm0.7\%$ 
                      & $89.7\pm0.6\%$ 
                      & $48.6\pm0.9\%$ 
                      & $57.4\pm0.6\%$ 
                      & $63.9\pm0.7\%$ 
                      & $67.8\pm0.5\%$ \\
      XgbTree & $35.4\pm1.5\%$ 
            & $58.4\pm1.9\%$ 
            & $74.2\pm1.5\%$ 
            & $88.7\pm1.1\%$ 
            & $49.9\pm1.6\%$ 
            & $57.8\pm1.3\%$ 
            & $64.1\pm1.0\%$
            & $68.4\pm0.8\%$ \\ 
      LambdaMART & $32.6\pm1.7\%$ 
                  & $56.2\pm2.2\%$ 
                  & $73.7\pm1.7\%$ 
                  & $88.3\pm0.8\%$ 
                  & $47.5\pm1.9\%$ 
                  & $56.3\pm1.7\%$ 
                  & $62.5\pm1.2\%$ 
                  & $67.1\pm1.0\%$ \\
      AdaRank & $29.9\pm1.3\%$ 
            & $53.3\pm1.1\%$ 
            & $71.4\pm0.9\%$ 
            & $85.8\pm0.8\%$ 
            & $44.7\pm1.1\%$ 
            & $53.7\pm0.8\%$ 
            & $59.9\pm0.8\%$
            & $65.4\pm0.6\%$ \\ 
      AnswerBot  & $31.8\pm1.8\%$ 
               & $52.8\pm2.0\%$ 
               & $71.6\pm1.9\%$ 
               & $88.7\pm0.9\%$ 
               & $44.5\pm1.7\%$ 
               & $54.3\pm1.6\%$ 
               & $62.6\pm1.2\%$ 
               & $68.1\pm0.9\%$ \\
      DeepAns-IR  & $42.8\pm1.3\%$ 
               & $62.8\pm1.9\%$ 
               & $78.2\pm2.0\%$ 
               & $90.0\pm0.9\%$ 
               & $55.4\pm1.6\%$ 
               & $63.1\pm1.6\%$ 
               & $68.2\pm1.0\%$ 
               & $72.1\pm0.8\%$ \\
      \hline            
      \textbf{DeepAns}  & $\mathbf{45.7\pm1.6\%}$ 
                     & $\mathbf{65.7\pm1.6\%}$ 
                     & $\mathbf{80.2\pm1.9\%}$ 
                     & $\mathbf{92.1\pm1.2\%}$ 
                     & $\mathbf{58.3\pm1.4\%}$ 
                     & $\mathbf{65.6\pm1.3\%}$
                     & $\mathbf{70.7\pm1.1\%}$
                     & $\mathbf{73.8\pm0.8\%}$ \\ 
      \hline
\end{tabular}
}
}
\end{table*}

\begin{table*}[htb]
\centering
\caption{\revv{Automatic evaluation (SO-Java)}}
\label{tab:automatic_eval_java}
\revv{
\resizebox{0.95\textwidth}{!}{
    \begin{tabular}{||l|cccc|cccc||} 
      \hline
      Model & P@1  & P@2 & P@3 & P@4 & DCG@2 & DCG@3 & DCG@4 & DCG@5 \\
      \hline
      RandomForest & $32.9\pm1.0\%$ 
                      & $56.1\pm1.1\%$ 
                      & $74.1\pm1.1\%$ 
                      & $89.5\pm0.7\%$ 
                      & $47.6\pm0.9\%$ 
                      & $56.6\pm0.8\%$ 
                      & $63.2\pm0.6\%$ 
                      & $67.2\pm0.5\%$ \\
      XgbTree & $35.9\pm1.3\%$ 
            & $59.0\pm1.2\%$ 
            & $75.7\pm0.9\%$ 
            & $89.1\pm1.0\%$ 
            & $50.5\pm1.0\%$ 
            & $58.8\pm0.7\%$ 
            & $64.6\pm0.7\%$
            & $68.8\pm0.5\%$ \\ 
      LambdaMART & $31.5\pm1.2\%$ 
                  & $54.4\pm1.2\%$ 
                  & $72.3\pm1.7\%$ 
                  & $87.6\pm1.3\%$ 
                  & $46.0\pm0.8\%$ 
                  & $54.9\pm1.1\%$ 
                  & $61.5\pm0.7\%$ 
                  & $66.3\pm0.5\%$ \\
      AdaRank & $29.2\pm2.1\%$ 
                  & $52.1\pm2.2\%$ 
                  & $69.9\pm1.8\%$ 
                  & $86.2\pm1.5\%$ 
                  & $43.6\pm1.8\%$ 
                  & $52.5\pm1.7\%$ 
                  & $59.6\pm1.5\%$ 
                  & $64.9\pm1.1\%$ \\
      AnswerBot  & $34.7\pm1.5\%$ 
               & $58.0\pm2.1\%$ 
               & $77.8\pm1.9\%$ 
               & $90.2\pm1.5\%$ 
               & $49.4\pm1.6\%$ 
               & $59.3\pm1.5\%$ 
               & $64.7\pm1.1\%$ 
               & $68.4\pm0.8\%$ \\
      DeepAns-IR  & $42.3\pm2.9\%$ 
               & $63.7\pm2.3\%$ 
               & $78.3\pm2.1\%$ 
               & $91.8\pm1.6\%$ 
               & $55.7\pm2.4\%$ 
               & $63.1\pm2.2\%$ 
               & $68.9\pm1.8\%$ 
               & $72.1\pm1.4\%$ \\
      \hline
      \textbf{DeepAns}  & $\mathbf{45.5\pm1.6\%}$ 
                     & $\mathbf{65.9\pm2.2\%}$ 
                     & $\mathbf{79.9\pm1.6\%}$ 
                     & $\mathbf{92.0\pm0.9\%}$ 
                     & $\mathbf{58.4\pm1.9\%}$ 
                     & $\mathbf{65.4\pm1.5\%}$
                     & $\mathbf{70.6\pm1.2\%}$
                     & $\mathbf{73.7\pm0.9\%}$ \\ 
      \hline
\end{tabular}
}
}
\end{table*}

    \item Compared to our model, the learning-to-rank based approach achieved the worst performance regarding the $P@K$ and $DCG@K$ scores with different depths. The learning to rank approach ignores the fact that ranking is a prediction task on a list of objects. Because they require a large number of training instances with ranking labels, therefore if the ground truth ordering of input candidates is lacking, they are unable to capture the relative preference between two QA pairs. This may explain the reason why its performance is comparatively suboptimal.
    
    \revv{
    \item The DeepAns-IR approach has its advantage as compared to other baselines excluding our proposed model. This is because DeepAns-IR employs the same data labeling strategy and the model structure as ours. Moreover, it also incorporates the IR-based approach to expand the query with clarifying questions. This verifies the effectiveness of our model for question and answering tasks in technical Q\&A sites. The only difference between DeepAns-IR and our model is that our model generates clarifying questions via deep sequence to sequence learning, while the DeepAns-IR retrieves the clarifying questions from the existing database according to a similarity score, which relies heavily on whether similar questions can be found and how similar the questions are. This results in our model's superior performance as compared to the DeepAns-IR approach. 
    }
    \item  \revv{By comparing the evaluation results of the different technical Q\&A sites, i.e., Ask Ubuntu, Super User and Stack Overflow, we can see that our proposed model is stably and substantially better than the other baselines.
    This suggests that our approach behaves consistently across different technical Q\&A platforms, regardless of the different topic of the specific technical forums. This supports the likely generalization and robustness of our approach. We also notice that the advantage of our proposed model is much more obvious on SO (Python) and SO (Java) as compared to Ask Ubuntu and Super User. 
    The reason for this phenomenon is likely the large number of training samples from Stack Overflow which benefits the classification performance of our model.
    }
\end{enumerate}


\revv{
In summary, our model substantially outperforms the baselines under automatic evaluation.
}
\subsection{\textbf{RQ-2: Ablation Analysis}}
Ablation analysis is used to verify the effectiveness of the \emph{DeepAns} using \emph{Question boosting} and \emph{Label establishing} methods. More specificly, we compare our approach with several of its incomplete variants:
\begin{itemize}
    \item \textbf{Drop CQ}: removes the clarifying question part generated by \emph{Question boosting} model.
    \item \textbf{Drop Labeling}: removes the training samples generated by \emph{Label establishing} model, to do this, we keep the best QA pairs as \emph{positive samples}, and make other answer pairs as \emph{negative samples}. Our model was trained as a binary classification model.
\end{itemize}
\revv{We performed the ablation analysis experiment on Ask Ubuntu and Super User respectively}. The ablation analysis results are presented in the Table~\ref{tab:ab_eval_ask} and Table~\ref{tab:ab_eval_super}. We can observe the following points.

\begin{table}
\caption{Ablation Evaluation (Ask Ubuntu)}
\label{tab:ab_eval_ask}
\begin{center}
\begin{tabular}{|c|c|c|c|}
    \hline
    {\bf Measure} & {\bf Drop CQ} & {\bf Drop Labeling} & {\bf DeepAns} \\
    \hline\hline
    P@1  & $34.2\pm1.3\%$ & $31.3\pm1.2\%$ & $\mathbf{40.9\pm1.5\%}$ \\
    \hline
    P@2 & $58.9\pm1.8\%$ & $50.5\pm1.1\%$  & $\mathbf{61.7\pm1.9\%}$ \\
    \hline
    P@3 & $77.3\pm1.5\%$ & $68.9\pm1.3\%$ & $\mathbf{77.9\pm0.9\%}$ \\
    \hline
    P@4 & $91.4\pm0.8\%$ & $86.0\pm1.1\%$ & $\mathbf{92.0\pm0.9\%}$ \\
    \hline
    DCG@2 & $49.8\pm1.5\%$ & $43.4\pm0.9\%$ & $\mathbf{54.0\pm1.7\%}$ \\
    \hline
    DCG@3 & $59.5\pm1.2\%$  & $51.7\pm0.8\%$ & $\mathbf{62.1\pm1.1\%}$ \\
    \hline
    DCG@4 & $65.8\pm0.9\%$ & $59.1\pm0.7\%$ & $\mathbf{68.2\pm1.1\%}$ \\
    \hline
    DCG@5 & $68.5\pm0.7\%$ & $64.5\pm0.5\%$ & $\mathbf{71.3\pm0.9\%}$  \\
    \hline
\end{tabular}
\end{center}
\end{table}

\begin{table}
\caption{Ablation Evaluation (Super User)}
\label{tab:ab_eval_super}
\begin{center}
\begin{tabular}{|c|c|c|c|}
    \hline
    {\bf Measure} & {\bf Drop CQ} & {\bf Drop Labeling} & {\bf DeepAns} \\
    \hline\hline
    P@1  & $35.8\pm1.4\%$ & $29.7\pm1.4\%$ & $\mathbf{40.7\pm1.9\%}$ \\
    \hline
    P@2 & $60.2\pm1.0\%$ & $53.9\pm1.9\%$  & $\mathbf{65.8\pm1.1\%}$ \\
    \hline
    P@3 & $79.6\pm0.9\%$ & $72.5\pm1.5\%$ & $\mathbf{82.2\pm1.1\%}$ \\
    \hline
    P@4 & $92.1\pm0.5\%$ & $89.5\pm0.9\%$ & $\mathbf{93.9\pm0.8\%}$ \\
    \hline
    DCG@2 & $51.2\pm1.1\%$ & $45.0\pm1.6\%$ & $\mathbf{56.5\pm1.2\%}$ \\
    \hline
    DCG@3 & $60.9\pm0.9\%$ & $54.0\pm1.4\%$ & $\mathbf{64.7\pm1.2\%}$ \\
    \hline
    DCG@4 & $66.3\pm0.7\%$ & $61.4\pm1.1\%$ & $\mathbf{69.8\pm1.0\%}$ \\
    \hline
    DCG@5 & $69.3\pm0.7\%$ & $65.6\pm0.8\%$ & $\mathbf{72.1\pm0.8\%}$  \\
    \hline
\end{tabular}
\end{center}
\end{table}

\begin{enumerate}
    \item By comparing the results of our approach with each of the variant model, we can see that no matter which method we dropped, it does hurt the performance of our model. This verifies the importance and effectiveness of these three mechanisms.
    \item By comparing the results of \textbf{DeepAns} with \textbf{Drop CQ}, it is clear that incorporating a clarifying question improves the overall performance.  When adding a clarifying question to our model, the $P@k$ score is improved by 19.5\% and 13.9\% on Ask Ubuntu and Super User dataset respectively. We attribute this to that the useful clarifying question can reduce the lexical gap between answer and questions, which can make the information properly fused between them.
    \item By comparing the results of \textbf{DeepAns} with \textbf{Drop Labeling}, we can measure the performance improvements achieved due to the incorporation of ``Label establishment'' process. After removing the training samples constructed by \emph{Label establishment}, there is a significant drop overall in every evaluation measure. This is because by employing our \emph{label establishing} process, the size of the training data is largely expanded, in the meanwhile, by introducing \emph{neutral$^+$} and \emph{neutral$^-$} samples, our model can learn to better distinguish best answer from similar ones.
\end{enumerate}
In summary, both the \emph{question boosting} module and \emph{label establishing} model are effective and helpful to enhance the performance of our approach.

\subsection{\textbf{RQ-3: Parameters Tuning}}
\revv{In this section, we tune the key parameters of our model for sensitivity analysis and robustness analysis.}

\begin{figure}\vspace*{-0.0cm}
\centerline{\includegraphics[width=0.45\textwidth]{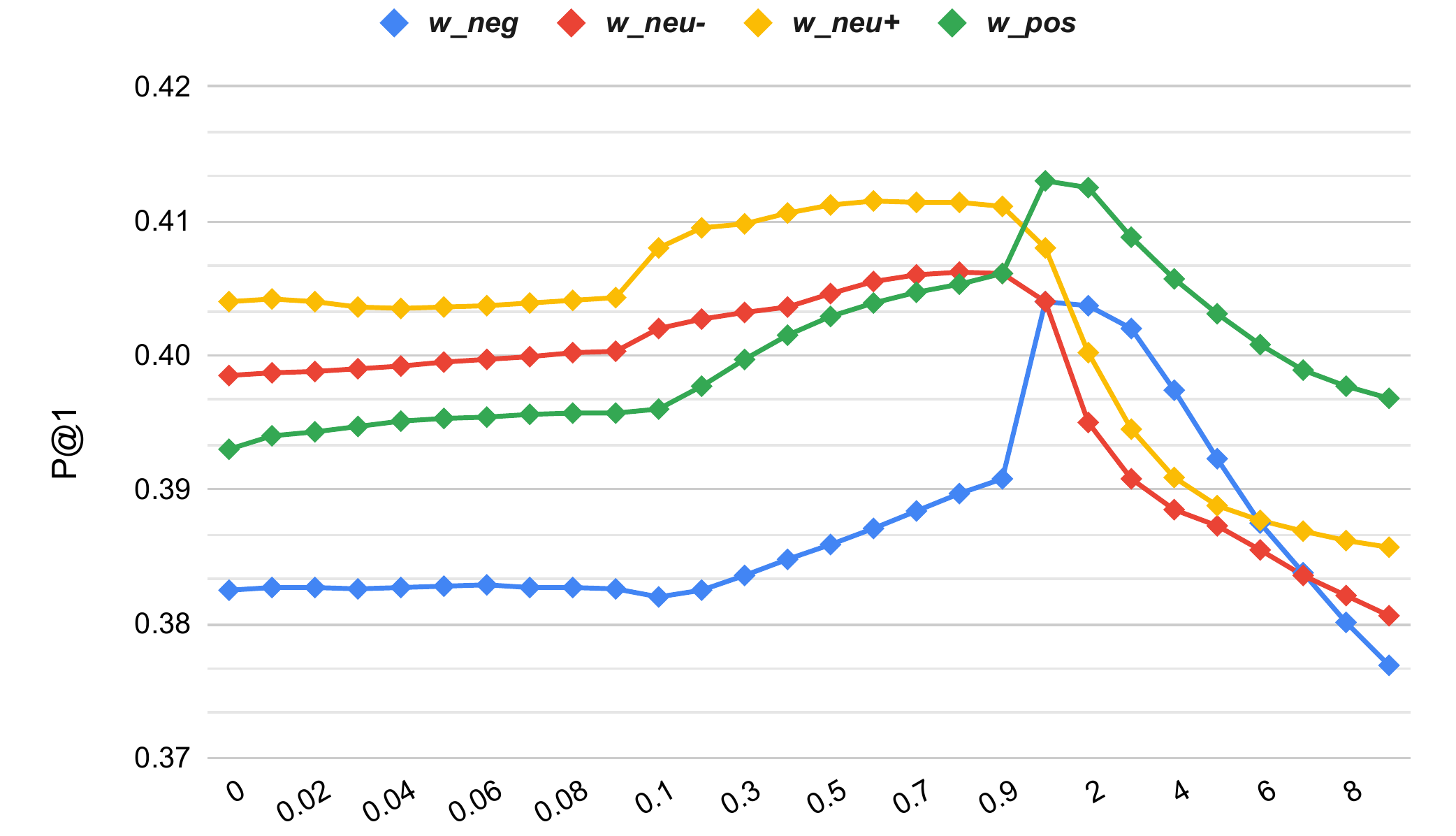}
			\includegraphics[width=0.45\textwidth]{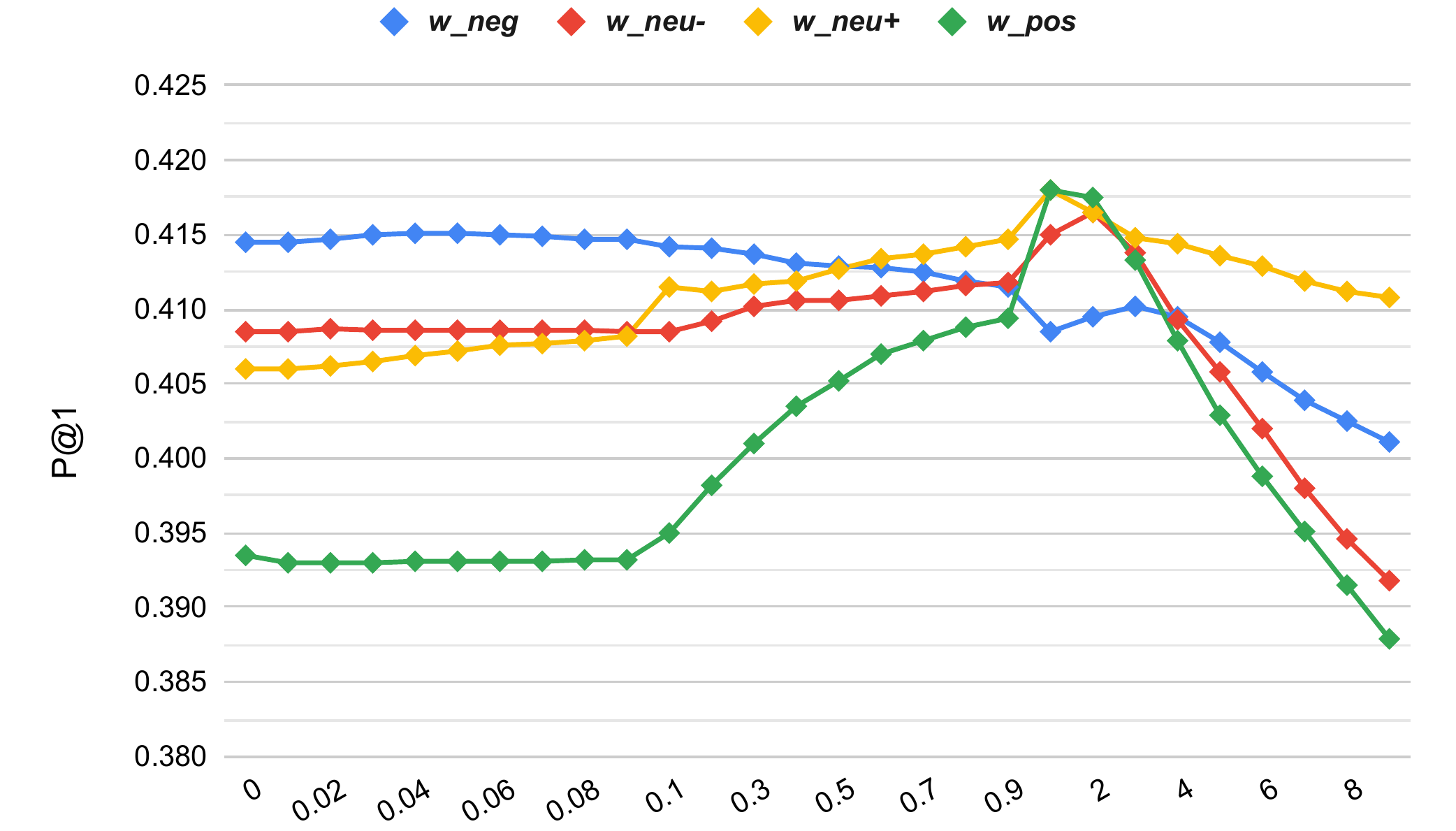}}
\caption{\revv{Sensitivity Analysis on Ask-ubuntu (left) and Super-user (right)}}
\label{fig:weighttuning}
\end{figure}

\subsubsection{\revv{Sensitivity Analysis}}
\label{subsec:ensitivity}
\revv{
We have four key parameters (i.e., $\omega_{pos}, \omega_{neu+}, \omega_{neu-}, \omega_{neg}$) in Equation~\ref{eq:score_eq}. The optimal settings of these weights were carefully tuned on our dataset. We demonstrate the weights tuning on Ask Ubuntu and Super User respectively.
In particular, the validation set was leveraged to validate our model and the grid search method was employed to select optimal parameters between $0$ and $10$ with small but adaptive step sizes. The step sizes were $0.01$, $0.1$, and $1$ for the range of [0, 0.1], [0.1, 1] and [1, 10], respectively. The parameters tuning process was varying one weight while fixing the other three weights. 
For example, in order to tune the parameter $\omega_{neg}$, we fix the other three parameters and change $\omega_{neg}$ from 0 to 10 with different step sizes. After that, we fix $\omega_{neg}$ to its optimal settings for tuning other parameters. Fig.~\ref{fig:weighttuning} illustrates the performance of our model with respect to different weights on Ask Ubuntu and Super User respectively. From the figure, we have the following observations:
\begin{enumerate}
    \item Even though the four parameters vary in a relatively wide range, the performance of our proposed model {\sc DeepAns} changes within small ranges near the optimal settings. This indicates that our model is non-sensitive to the parameters around their optimal settings, which further supports the generalization ability of our approach.
    \item We notice that most parameters achieve their best performance in the range of [1, 3], we thus recommend to initialize the weights in Equation~\ref{eq:score_eq} to be around the above range, which is close to the optimal settings of our model.
\end{enumerate}
}

\subsubsection{\revv{Robustness Analysis}}
In real world Q\&A sites, there is no guarantee to find the exactly matched questions from the archive, expecially when $k$ is small. Therefore we have to enlarge $k$ to improve the recall of the similar questions and hence the ``matched answers''. However, a larger $k$ may introduce more noise into the answer candidate pool with more irrelevant answers. This can then increase the difficulty of our answer recommendation task. 

\begin{figure}\vspace*{-0.0cm}
\centerline{\includegraphics[width=0.45\textwidth]{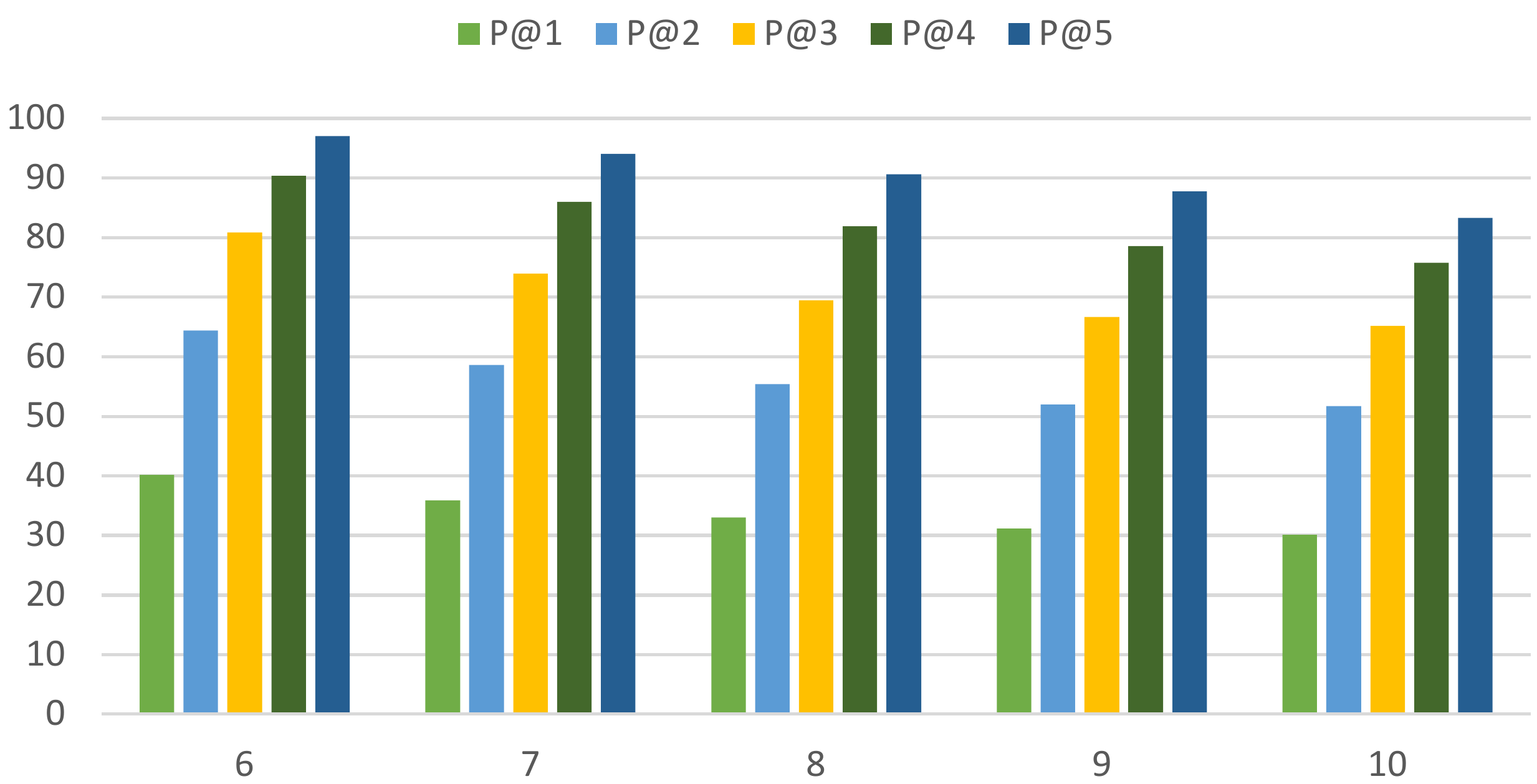}
			\includegraphics[width=0.45\textwidth]{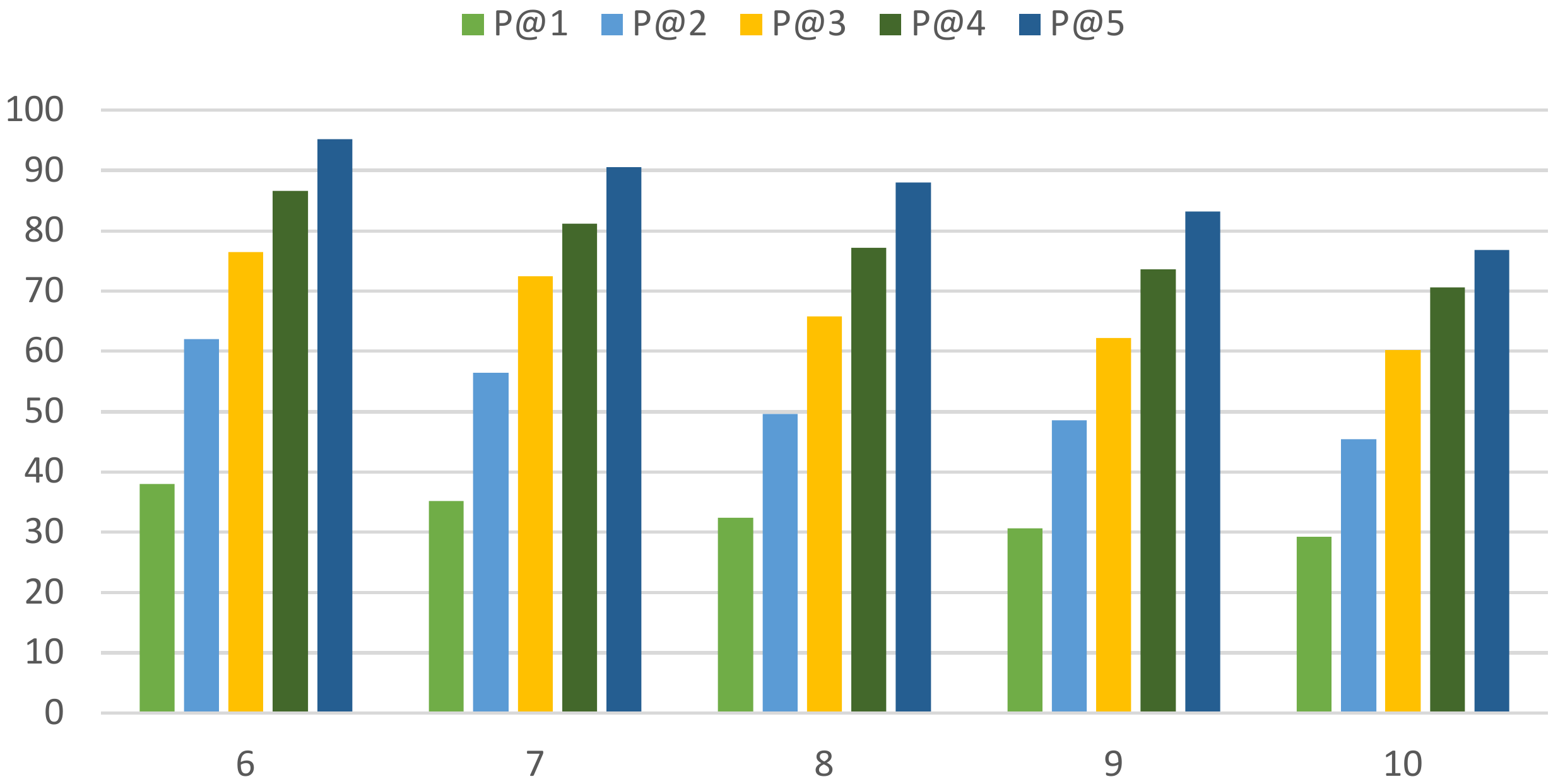}}
\caption{Robustness Analysis on Ask-ubuntu (left) and Super-user (right)}
\label{fig:parametertuning}
\end{figure}

To verify the robustness of our proposed approach, we set different thresholds for the number of returned questions by k-NN method. 
\minor{More specifically, we varied the number of returned similar questions $k$ from 6 to 10 and measured the performance of our approach, we then reported average P@1-5 over each dataset under different parameter settings of $k$.} 
The results \revv{of Ask Ubuntu and Super User} are shown in Fig.~\ref{fig:parametertuning}. We can make the following observations: 
\begin{enumerate}
\item The trend in overall performance of our model decrease as $k$ increases, which supports our concern that larger $k$ settings introduce more noises and bring bigger challenges for our task. 
\minor{By analyzing the performance of our approach with respect to different $k$, we notice that our approach achieves good performance when $k$ varies from 5 to 7, while still ensuring the ``matched answer'' is highly-ranked. We thus recommend setting $k$ within the above range for real-world applications.}
\item The advantage of our proposed model is more obvious on $P@1$ compared with other metrics($P@2-5$). Even when we set $k$ to 10, the performance of our model on $P@1$ is still on a par with the best performance of other baselines, while $k$ is set to 5 in these baselines (See Table~\ref{tab:automatic_eval_ask} and Table~\ref{tab:automatic_eval_super}). This reveals that our model can perform well under a noisy context, which shows the robustness of our model.
\end{enumerate}

In summary, our model is \revv{non-sensitive} and robust under different parameter settings.


%% file: human_results.tex
\revv{Since automatic evaluation results do not always agree with the actual ranking preference of real-world users, we also performed a small, qualitative user study to measure how humans actually perceive the results produced by our approach. Specifically, we mainly focus on the following research questions:
\begin{itemize}
    \item \textit{RQ-4:} How effective is the question boosting results of our approach under human evaluation? 
    \item \textit{RQ-5:} How effective is the question answering results of our approach under human evaluation?
\end{itemize}
}
\revv{
For human evaluation, we used the Ask Ubuntu and Stack Overflow (Python) platforms to perform our user study. We invited 5 evaluators to participate in our user study; all of these participates have more than three years of studying/working experience in software development process, have more than one year of experience using technical Q\&A sites, and are familiar with the Ubuntu system and Python programming languages. We did not limit the amount of time for evaluators to complete the user study.
}

\revv{
\subsection{RQ4: Human Evaluation on Question Boosting Results} 
To gain a deeper understanding of how the clarifying questions impact the results in our study, we conducted human evaluation studies to measure how humans perceive the question boosting results. To do this, we consider two modalities in our user study: \textit{Relevance} and \textit{Usefulness}. \textit{Relevance} measures how relevant the clarifying question is to the original question title. \textit{Usefulness} measures how useful the clarifying question is for adding missing information for the original post. 
We randomly sampled 25 $\langle q, a \rangle$ pairs from Ask Ubuntu and SO (Python) respectively. 
For each question, we provided two clarifying questions. One was generated by our approach, the other was generated by the IR-based approach, i.e., DeepAns-IR.
We also provided the accepted answer to the question as a reference. 
We asked the participants to manually rate the generated clarifying questions on a scale between 1 and 3 (1 = worst, 3 = best) across the above modalities. The volunteers were blinded as to which question title was generated by our approach.
}

\begin{table*} 
\centering
\caption{\revv{Human Evaluation of Question Boosting Results}}
\label{tab:human_qb}
\vspace*{-6pt}
\revv{
\resizebox{0.95\textwidth}{!}{
    \vspace{-0.2cm}\begin{tabular}{||l|l||c c c c || c c c c ||}
        \hline
        \ Data & Model  & Score(1)\textsubscript{R}
                 & Score(2)\textsubscript{R}
                 & Score(3)\textsubscript{R}
                 & Avg\textsubscript{R}
                 & Score(1)\textsubscript{U}
                 & Score(2)\textsubscript{U}
                 & Score(3)\textsubscript{U} 
                 & Avg\textsubscript{U}\\
        \hline
        \multirow{2}{*}{\bf{Ask Ubuntu}} & IR-based 
        & 21.6\% & 43.2\% & 35.2\% & 2.14 
        & 28.8\% & 34.4\% & 36.8\% & 2.08
        \\ \cline{2-10}
        & Ours 
        & 18.4\% & 32.8\% & 48.8\% & 2.30 
        & 22.4\% & 35.8\% & 42.4\% & 2.20
        \\ \hline
        \multirow{2}{*}{\bf{SO (Python)}} & IR-based 
        & 19.2\% & 36.0\% & 44.8\% & 2.26 
        & 26.4\% & 32.0\% & 41.6\% & 2.15
        \\ \cline{2-10}
        & Ours 
        & 17.6\% & 32.0\% & 50.4\% & 2.33 
        & 23.2\% & 29.6\% & 47.2\% & 2.24 \\
        \hline
    \end{tabular}
}
}
\end{table*}

\noindent\textbf{\revv{Evaluation Results.}} 
\revv{
We obtained 125 groups of scores from evaluators for Ask Ubuntu and SO (Python) respectively. Each group contains two pairs of scores, which were rated for clarifying questions produced by IR-based approach and ours. Each pair contains a score for the \textit{Relevance} modality and a score for \textit{Usefulness} modality. 
The score distribution and average score of \textit{Relevance} and \textit{Usefulness} across the two methods are presented in Table~\ref{tab:human_qb}. From the table, we can observe the following points:
\begin{enumerate}
    \item Our approach performs better than the IR-based approach on both modalities. We attribute this to the following reason: the IR-based approach relies heavily on whether similar clarifying questions can be retrieved from the existing $\langle q, cq \rangle$ dataset. Considering the complexity of the questions in technical Q\&A sites, there may exist only a few questions that are very similar to the given one, hence it is difficult to retrieve relevant clarifying questions from the training set. 
    \item Both the IR-based approach and our approach can produce relevant and useful clarifying questions for the given question. This further verifies the clarifying question is helpful in adding missing information and reducing the gap between questions and answers. 
    We also notice that there are still quite a few questions that received low scores for \textit{Relevance} and \textit{Usefulness} modalities. Even though the clarifying questions generated by our approach are still not perfect, our study is the first step on this topic and we also release our data to inspire follow-up work for utilizing the clarifying questions. 
\end{enumerate}
}

\begin{figure} 
\centerline{\includegraphics[width=0.95\textwidth]{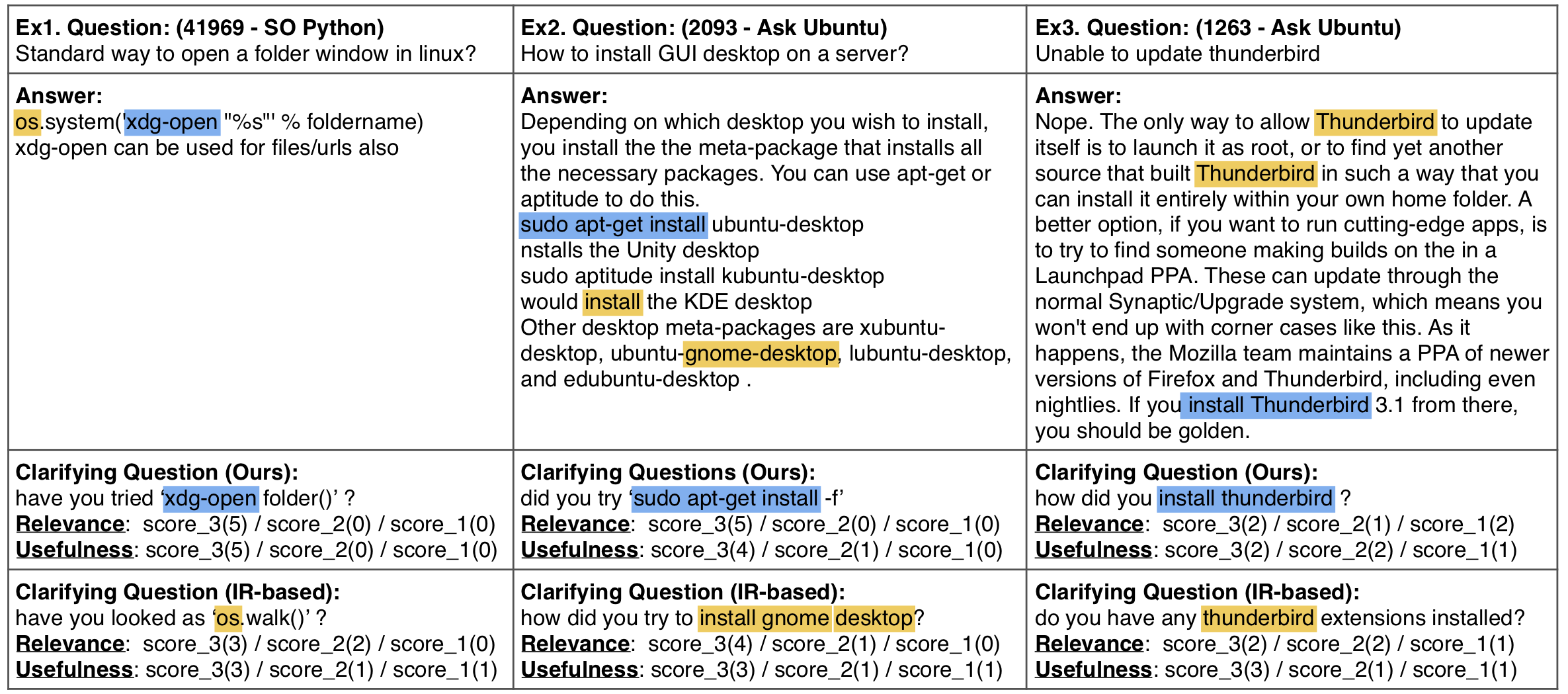}}
\caption{\revv{Evaluation Examples of Question Boosting.}}
\label{fig:example_qb}
\end{figure}

\noindent\textbf{\revv{Evaluation Examples.}} 
\revv{
A major challenge for question answering tasks is the semantic gap between the questions and answers. This is because the questions from technical Q\&A sites are, more often than not, very specific and complex, and oriented towards expert professional answers. To fill the gap between question and answers, we employ a deep encoder-decoder model to generate a clarifying question for a given post as a way of question boosting.
Fig.~\ref{fig:example_qb} presents three examples of human evaluation on question boosting results (the words that appear in both clarifying questions and answers are highlighted). From these cases, we can see that: 
\begin{enumerate}
    \item The clarifying questions produced by our approach as well as the IR-based approach generally perform well across both modalities. 
    It is clear that the clarifying question can reduce the lexical gap between the answer and the questions, which can add missing information and make the information better linked between question and answers. For example, in the first and second case, our approach generates ``xdg-open'' and ``sudo apt-get install'' for the clarifying questions which also appear in the answers. Thus, the added information can eliminate and/or reduce the isolation between questions and answers.
    We attribute this to the advantage of our model for learning common patterns automatically from the $\langle q, cq \rangle$ pairs.
    \item Not all the clarifying questions are appreciated by the evaluators; an example is shown in the last row of Fig.~\ref{fig:example_qb}. 
    For such cases, even though the generated clarifying question is not optimal to the participants, 
    our approach still precisely replicates the salient tokens, i.e., ``thunderbird'' from the question title, which also increases the likelihood of selecting the right answer from answer candidates.
\end{enumerate}
In summary, the clarifying questions generated by our approach are effective under human evaluation results.
}

\revv{
\subsection{RQ5: Human Evaluation on Question Answering Results}
Since the final goal of our study is recommending relevant answers to developers, we also performed a human evaluation to measure the effectiveness of question answering results with respect to human developers. 
\minor{To be more specific, we measured how developers perceive the answers produced by our approach to solved questions, unresolved questions and unanswered questions.
For solved questions, we compared our approach with the ground truth; for unresolved questions, we compared our approach with xgbTree and Answerbot methods; and for unanswered questions, we compared our approach with Stack Exchange search engine and Google search engine. 
} 
}

\revv{
\subsubsection{User Study on Solved Questions}
In order to investigate the agreement of the developers on solved questions, we randomly sampled 25 examples of solved questions from the testing set of Ask Ubuntu and SO (Python) respectively. For each solved question, we provided two answer candidates. One answer was the accepted answer -- we refer to it as the ground truth in this study. The other answer was produced by our approach. 
After that, each evaluator was asked to manually rate on the two answer candidates from 1 to 3, according to the acceptance of the answer. Score 3 means that the evaluator strongly agrees with the acceptance of the answer, and score 0 means that the evaluator strongly disagrees with the acceptance of the answer. 
It is worth emphasizing that the answer selected by our approach may actually be the same with the ground truth answer, and the participants were blinded as to which answer is the ground truth. 
}

\begin{table}
\caption{\revv{Human Evaluation - Ask Ubuntu}}
\label{tab:human_ubuntun}
\begin{center}
{\scriptsize
\revv{
\begin{tabular}{||l|c|c|c|c|c||}
\hline
    {\bf Type} & {\bf Approach} & {\bf Score(1)} & {\bf Score(2)} & {\bf Score(3)} & {\textbf{$Rank_{avg}$} }\\
    \hline\hline
    \multirow{2}{*}{\bf \revv{Solved}}
    & Ground Truth & 7.2\% & 18.4\% & 74.4\% & 2.67 \\ \cline{2-6}
    & DeepAns  & 19.2\% & 32.8\% & 48.0\% & 2.29 \\ \hline
    \multirow{3}{*}{\bf Unresolved}   
    & xgbTree       & 15.2\% & 26.4\% & 58.4\% & 2.43 \\ \cline{2-6}
    & AnswerBot     & 12.8\% & 28.0\% & 59.2\% & 2.46 \\ \cline{2-6}
    & DeepAns       & 12.0\% & 23.2\% & 64.8\% & 2.53 \\ \hline
    \multirow{3}{*}{\bf Unanswered}   
    & SE Engine     & 51.2\% & 33.6\% & 15.2\% & 1.64 \\ \cline{2-6}
    & Google        & 25.6\% & 32.8\% & 41.6\% & 2.16 \\ \cline{2-6}
    & DeepAns       & 22.4\% & 30.4\% & 47.2\% & 2.25 \\ \hline
\end{tabular}
}}
\end{center}
\end{table}

\begin{table}
\caption{\revv{Human Evaluation - SO (Python)}}
\label{tab:human_so}
\begin{center}
{\scriptsize
\revv{
\begin{tabular}{||l|c|c|c|c|c||}
\hline
    {\bf Type} & {\bf Approach} & {\bf Score(1)} & {\bf Score(2)} & {\bf Score(3)} & {\textbf{$Rank_{avg}$} }\\
    \hline\hline
    \multirow{2}{*}{\bf \revv{Solved}}
    & Ground Truth & 5.6\% & 14.4\% & 80.0\% & 2.74 \\ \cline{2-6}
    & DeepAns  & 18.4\% & 31.2\% & 50.4\% & 2.32 \\ \hline
    \multirow{3}{*}{\bf Unresolved}   
    & xgbTree   & 12.0\% & 26.4\% & 61.6\% & 2.50 \\ \cline{2-6}
    & AnswerBot & 9.6\%  & 32.0\% & 58.4\% & 2.49 \\ \cline{2-6}
    & DeepAns   & 10.4\% & 21.6\% & 68.0\% & 2.58 \\ \hline
    \multirow{3}{*}{\bf Unanswered}   
    & SE Engine  & 54.4\% & 32.0\% & 13.6\% & 1.59 \\ \cline{2-6}
    & Google     & 30.4\% & 31.2\% & 38.4\% & 2.08 \\ \cline{2-6}
    & DeepAns    & 26.4\% & 28.0\% & 45.6\% & 2.19 \\ \hline
\end{tabular}
}}
\end{center}
\end{table}

\begin{figure} 
\centerline{\includegraphics[width=0.95\textwidth]{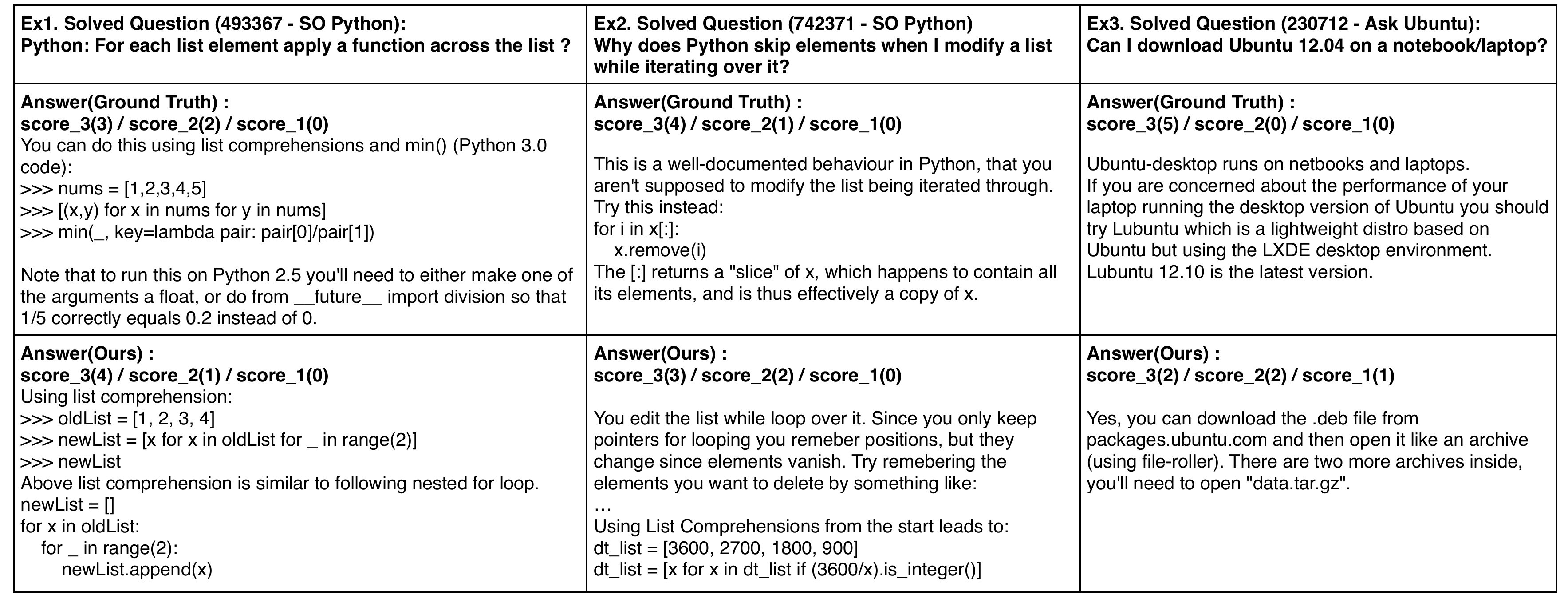}}
\caption{\revv{Examples of Solved Questions}}
\label{fig:example_solved}
\end{figure}

\vspace{0.1cm}\noindent\textbf{\revv{Evaluation Results.}} 
\revv{
We collected 125 groups of scores from participants for Ask Ubuntu and SO (Python) respectively. Each group contains two scores, which were rated for answers of the ground truth and ours. We count the proportion of different scores and calculate the average score for each method. The evaluation results for Ask Ubuntu and SO (Python) are presented in Table~\ref{tab:human_ubuntun} and Table~\ref{tab:human_so} respectively. From the table, we can observe the following points:
\begin{enumerate}
    \item The evaluators are in agreement with acceptance of the ground truth answers for most cases. For example, around 75\% of the ground truth answers in Ask Ubuntu and  80\% answers in SO (Python) are appreciated by the volunteers. 
    \item The ground truths are better than our approach. This is reasonable because the ground truth answers are usually high-quality answers that have been accepted by the developers. Even though our approach is not as good as the ground truth at the current stage, we observe that a small number of answers produced by our approach are marked with score 1. This indicates that the answers selected by our approach are meaningful and acceptable for the majority of questions. 
\end{enumerate}
}
\vspace{0.1cm}\noindent\textbf{\revv{Evaluation Examples.}} 
\revv{
Fig.~\ref{fig:example_solved} shows three examples of the user study on solved questions. It can be seen that:
\begin{enumerate}
    \item In general, our approach can produce acceptable answers. Sometimes, the answers chosen by our approach are actually more accepted by the volunteers than the ground truth answers. For example, in the first sample, three evaluators gave a score of 3 to the ground truth answer, while four evaluators gave a score of 3 to ours. 
    However, our answer does not belong to the current question thread and is selected from answer candidates of other questions (e.g., \textit{Python: duplicating each element in a list}). This further justifies the feasibility of addressing \emph{answer hungry} problem by selecting answers from the historical QA dataset.
    \item Outputs from our model are not always ``correct''. For example, in the last sample, the information seeker asks a question of ``\textit{Can I download Ubuntu 12.04 on a notebook/laptop?}'', while the answer provided by our approach is about how to download a file from the packages. This example reveals that considering the complexity of the questions in technical Q\&A sites, the gap between the ground truth answers and ours is still large, and hence there is still a large room for our question answering system to be further improved.
\end{enumerate}
}

\subsubsection{\revv{User Study on Unresolved Questions}}
\revv{To investigate how developers perceive our approach to solve the unresolved questions, we sampled 25 unresolved questions for Ask Ubuntu and SO (Python) respectively. Each question has multiple answer candidates that have not been selected as \emph{Accept}. By computing the matching score between question and each answer candidate, we can identify a best answer via our approach, xgbTree and Answerbot respectively (note that different approaches may choose the same answer as the best answer). Following that, we ask each evaluator to rank three answer candidates produced by our approach, xgbTree and Answerbot from 1 to 3 (3 is the best) according to the acceptance of the answer. It is worth emphasizing that the answers identified by our approach and others could be the same, and the order of the answers is randomly decided.}

\vspace{0.1cm}\noindent\textbf{\revv{Evaluation Results.}} 
\revv{
The human evaluation results of unresolved questions for Ask Ubuntu and SO (Python) are presented in Table~\ref{tab:human_ubuntun} and Table~\ref{tab:human_so} respectively. From the table, we can see that:
\begin{enumerate}
    \item Our model performs better than xgbTree and Answerbot baselines. This further indicates that the answers selected by our approach are more appreciated by evaluators. The results of human evaluation on unresolved questions are consistent with large-scale automatic evaluation results, which reconfirms the effectiveness of our approach for identifying the best answer in unresolved questions.
    \item Compared with the evaluation results of ground truth, the average scores between the answers of unresolved questions and solved questions are close, which supports our previous assumption that users forget to mark the accepted answer is not uncommon in technical Q\&A sites.
\end{enumerate}
}

\begin{figure} 
\centerline{\includegraphics[width=0.95\textwidth]{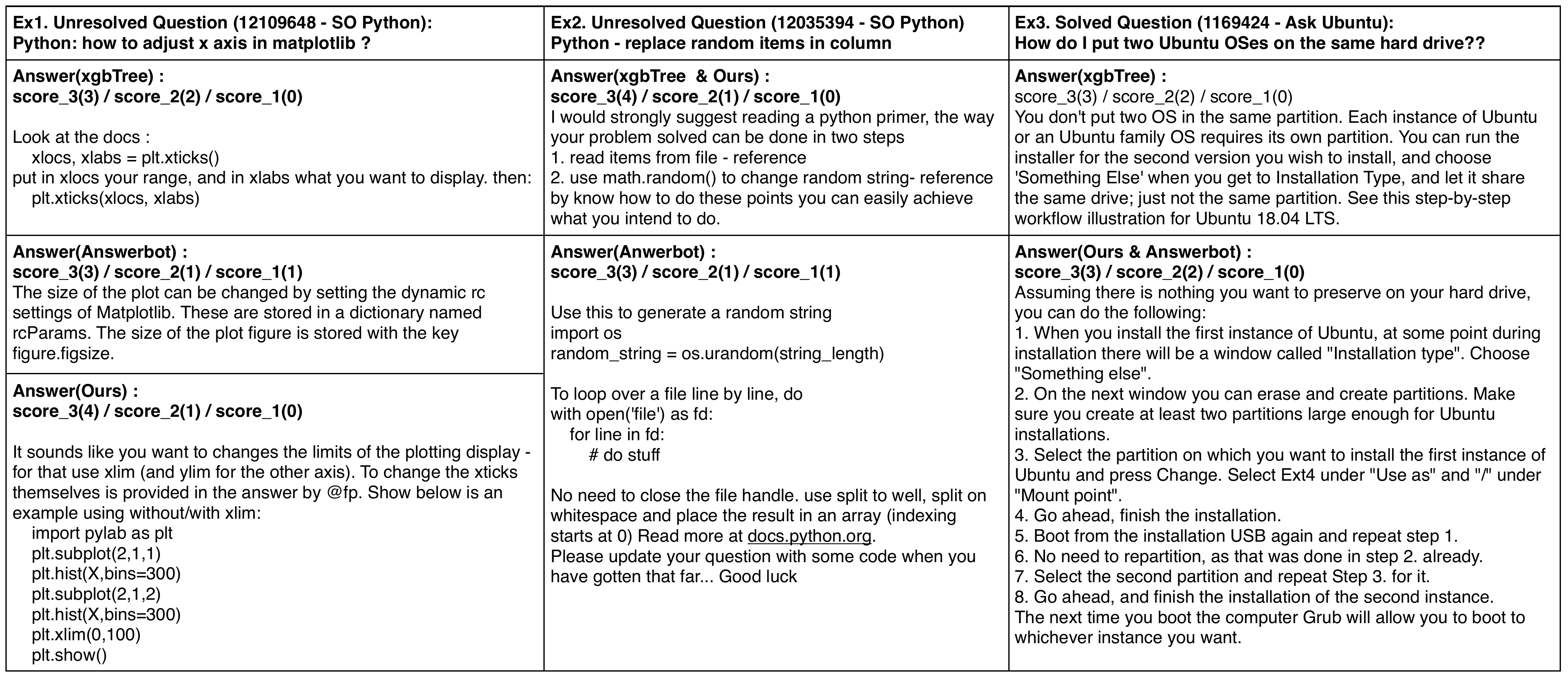}}
\caption{\revv{Examples of Unresolved Questions}}
\label{fig:example_unresolved}
\end{figure}

\vspace{0.1cm}\noindent\textbf{\revv{Evaluation Examples.}} 
\revv{
Fig.~\ref{fig:example_unresolved} shows the examples of the user study on unresolved questions. It can be seen that: 
\begin{enumerate}
    \item The overall answer quality for the unresolved questions is good. This is because these answers are directly related to the specific problems of the questions, which are more suitable to the needs of information seekers.
    \item Even all the answer candidates of an unresolved question aim at solving the same problem. As can be seen, some answers identified by our approach stand out from the rest and are more appreciated by evaluators, such as samples 1-2. This further verifies the ability of our approach to select the most relevant answer from a set of answer candidates.
\end{enumerate}
}

\subsubsection{\revv{User Study on Unanswered Questions}}
\revv{
Similar to unresolved questions, We also randomly sampled 25 examples of unanswered questions for Ask Ubuntu and SO(Python) respectively. 
For each unanswered question, considering that developers usually search for technical help using Google search engine and/or the Q\&A site search engine itself, we compare our approach against two baselines built based on the above search engines respectively. 
We used the question title of the post as the search query. 
For Google search engine, we add ``site:stackoverflow.com'' and ``site:askubuntu.com'' to the end of the search query so that it searches only posts on Stack Overflow and Ask Ubuntu respectively. We use the first ranked question returned by Google search engine as the most relevant question, we extracted the accept answer or the answer with the highest vote if there is no accepted answer of the relevant question. 
For technical Q\&A site search engine, we refer to the first ranked related question recommended by the technical Q\&A site search engine as the most relevant question, and extracted the associated accepted answer or the highest-vote answer. After constructing the evaluation set for unanswered questions, for each unanswered question, we asked the evaluators to rank on the 3 answer candidates from 1 to 3 (3 for the best answer), The higher grade indicates that the answer is more suitable to the given question. Please note that the participants do not know which answer is generated by which approach.
}

\vspace{0.1cm}\noindent\textbf{\revv{Evaluation Results.}}
\revv{
The expert evaluation results of unanswered questions for Ask Ubuntu and SO (Python) are presented in Table~\ref{tab:human_ubuntun} and Table~\ref{tab:human_so}. We can observe the following points:
\begin{enumerate}
    \item Compared with baselines, our model outperforms SE (Stack Exchange search engine) and Google (Google search engine). 
    This suggests that the answers produced by our approach are considered to be more suitable to the given question by the evaluators. 
    We attribute this to the reason that Google search engine identifies the answer via searching from similar questions, thus it is unable to judge the matching degree between the questions and answers. In contrast, our approach estimates the matching score using the context information of the qa pair, which fills the gap between questions and answers.
    The superior performance of our approach in terms of average score further supports the effectiveness of our approach in identifying the best answer.
    \item For the unanswered questions, a gap for the answer quality between unanswered questions and solved/unresolved questions still exists. We also notice that our approach received more low scores ($score=1$) with unanswered questions as compared to solved/unresolved questions. 
    This is because in technical Q\&A sites, some questions are rather complicated and sophisticated and it is hard to find suitable question-specific answers for these questions.
\end{enumerate}
}

\begin{figure} 
\centerline{\includegraphics[width=0.95\textwidth]{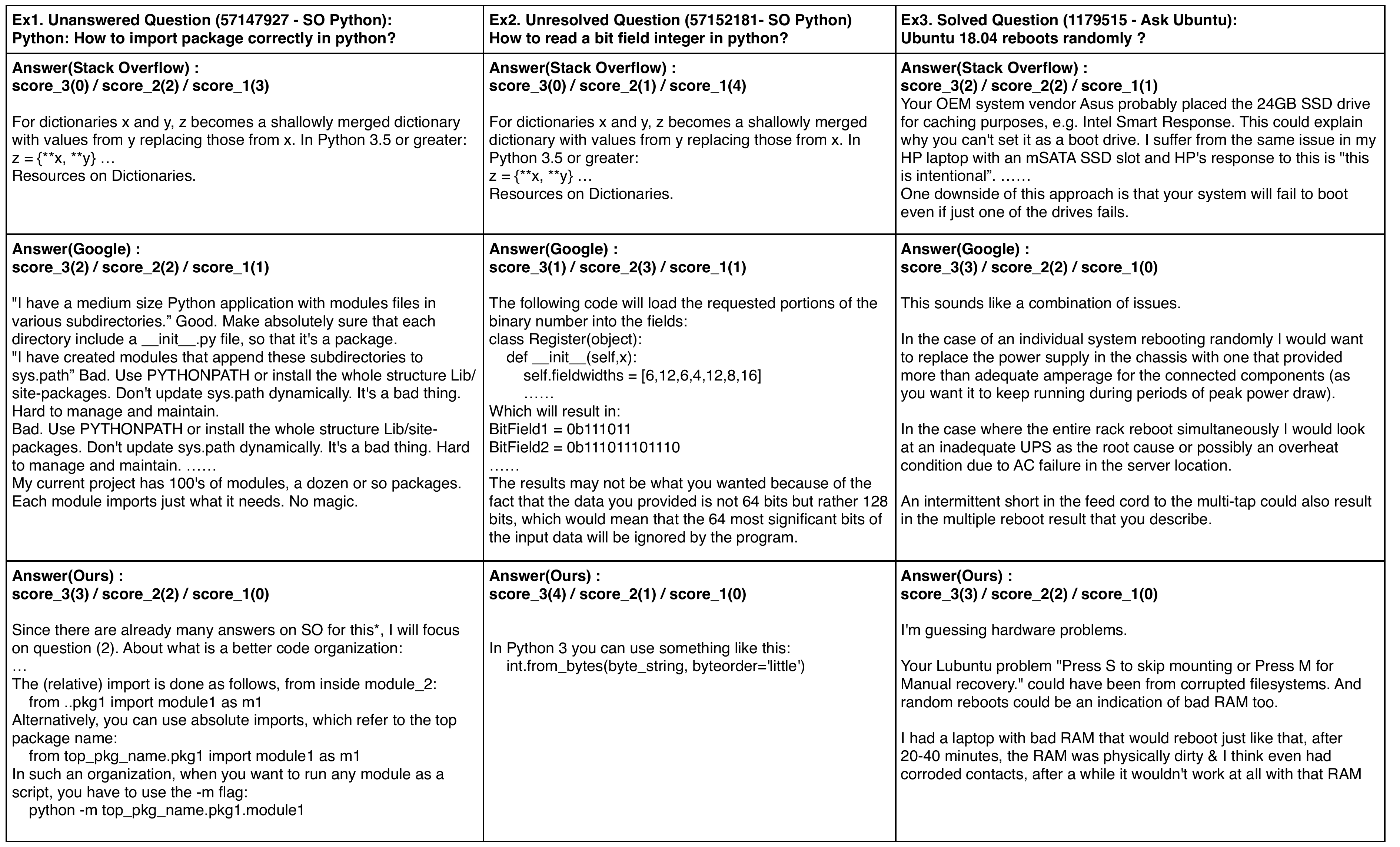}}
\caption{\revv{Examples of Unanswered Questions}}
\label{fig:example_unanswered}
\end{figure}

\vspace{0.1cm}\noindent\textbf{\revv{Evaluation Examples.}} 
\revv{
Fig.~\ref{fig:example_unanswered} shows three examples of the user study on unanswered questions. we can observe the following points:
\begin{enumerate}
    \item The search engine of the technical Q\&A site achieves worst performance. For example, in sample 1 and sample 2, the SE search engine  recommends the same answer to two different questions. This is why the evaluators give comparatively low scores to the answers identified by SE search engine.
    \item Our approach has its advantage as compared to the Google search engine (e.g., sample 1-2).  
    This is because the Google search engine does not consider the contextual information between the questions and answers, but instead only identifies the answers based solely by searching for similar questions. By contrast, our approach takes the question as well as the candidate answers and calculates the matching score between the question and the answers, which results in its superior performance compared to the other baselines.
    \item In technical Q\&A sites, some question titles are relatively abstract and uninformative. For example, in sample 3, even the answer selected by our approach is relevant and meaningful, we can not make sure if the answer solves the actual problem or not. For such cases, more detailed information, such as the description in the question body, could be considered when searching for appropriate answers.  
\end{enumerate}
}

\revv{
In summary, our model is comparatively effective under human evaluation for question answering tasks in technical Q\&A sites.
}

%% file: disc.tex
\revv{
In this section, we first discuss the strength of our approach as well as the  threats to validity of our work, after that we analyze the outlier cases involving in our data creation process. 
}

\subsection{\revv{Strength of Our Approach}}
\revv{To address the answer hungry problem in technical Q\&A sites, we propose a deep learning based approach {\sc DeepAns} to search relevant answers from historical QA pairs. We summarized the strength of our approach as follows:
}

\subsubsection{\revv{Neural Language Model for Question Boosting}}
\revv{
One advantage of our approach is training an attentional sequence-to-sequence model for generating clarifying questions as a way of question boosting. Instead of searching similar clarifying questions, our approach builds a neural language model for linking semantics of question and clarifying questions. 
The neural language model is able to handle the uncertainty in the correspondence between the questions and clarifying questions. Our approach automatically learns common patterns automatically from the $\langle q, cq \rangle$ pairs. 
The encoder itself is a neural language model which is able to remember the likelihood of different kinds of questions. Following that, the decoder learns the context of the questions fills the gap between the questions and clarifying questions.
}
\subsubsection{\revv{Label Establishment for Data Augmentation}} 
\revv{ 
Due to the reason of the professional questions in technical Q\&A sites, it is thus very hard, if not possible, to find experts and annotators for manual labeling the QA pairs. In this paper, we present a novel labeling scheme to automatically construct \emph{positive}, \emph{neutral$^+$}, \emph{neutral$^-$}, and \emph{negative} training samples. Guided by our four heuristic rules, this label establishment process can collect large amounts of labeled QA pairs, which greatly saves the time-consuming and labor-intensive labeling process. 
}

\subsubsection{\revv{Deep Neural Network for Answer Recommendation}}
\revv{
We present a weakly supervised neural network for the answer recommendation task in technical Q\&A sites. Our model architecture is able to incorporate the aforementioned four types of training samples for ranking QA pairs. Our work first uses the deep neural network to solve the problem of best answer selection in technical Q\&A sites, which is able to alleviate the answer hungry phenomenon that widely exists in technical Q\&A forums.
}

\subsection{Threats to Validity}
\label{sec:threats}
\revv{We have identified the following threats to validity among our study:}

\vspace{0.1cm}\noindent \textbf{\revv{Internal Validity}}
\revv{
Threats to internal validity are concerned
with potential errors in our code implementation and study settings. For the automatic evaluation, in order to reduce errors, we have double-checked and fully tested our source code. We have carefully tuned the parameters of the baseline approaches and used them in their highest performing settings for comparison, but there may still exist errors that we did not note. Considering such cases, we have published our source code and dataset to facilitate other researchers to replicate and extend our work.
}

\vspace{0.1cm}\noindent \textbf{\revv{External Validity}}
\revv{
The external validity relates to the quality and generalizability of our dataset. Our dataset is constructed from the official StackExchange data dump. We focus on three technical Q\&A sites, i.e., Ask Ubuntu, Super User and Stack Overflow for our experiment. These three technical Q\&A sites are commonly used by software developers and each one focuses on a specific area.  
However, there are still many other technical Q\&A sites in StackExchange which are not considered in our study (e.g., Server Fault).  
We believe that our results will generalize to other technical Q\&A sites as well, due to the ability of our approach to identify the best answer from a set of answer candidates. We will try to extend our approach to other technical Q\&A sites to benefit more users in future studies.
}

\vspace{0.1cm}\noindent \textbf{\revv{Construct Validity}}
\revv{
The construct validity concerns the relation between theory and observation. In this study, such threats are mainly due to the suitability of our evaluation measures.
For human evaluation, the subjectiveness of the evaluators, the evaluators' degree of carefulness, and the human errors may affect the validity of judgements.
We minimized such threats by choosing experienced participants who have at least three years of studying/working experience in the software development process, and are familiar with Ubuntu system and Python programming languages. We also gave the participants enough time to complete the evaluation tasks.
}

\vspace{0.1cm}\noindent\textbf{\revv{Model Validity}}
\revv{
The model validity relates to model structure that could affect the learning performance of our approach. 
In this study, for the answer recommendation task, we choose a CNN-based model due to the optimum results achieved by ~\cite{kim2014convolutional}. 
Recent studies~\cite{lai2015recurrent, zhou2016text} have shown that the RNN-based model can also achieve promising performance on the text classification task, which is similar to ours. 
For the question boosting task, we use the vanilla sequence-to-sequence model. Recent research has proposed new models, such as the pointer-generator~\cite{see2017get}, transformer~\cite{vaswani2017attention} and bert~\cite{devlin2018bert}. 
However, our results do not shed light on the effectiveness of  employing other deep learning models with respect to different structures and new advanced features. 
We will try to use other deep learning models for our tasks in future work and compare them to those we report in this paper.
}




\begin{figure}\vspace{-0.1cm}
\centerline{\includegraphics[width=0.85\textwidth]{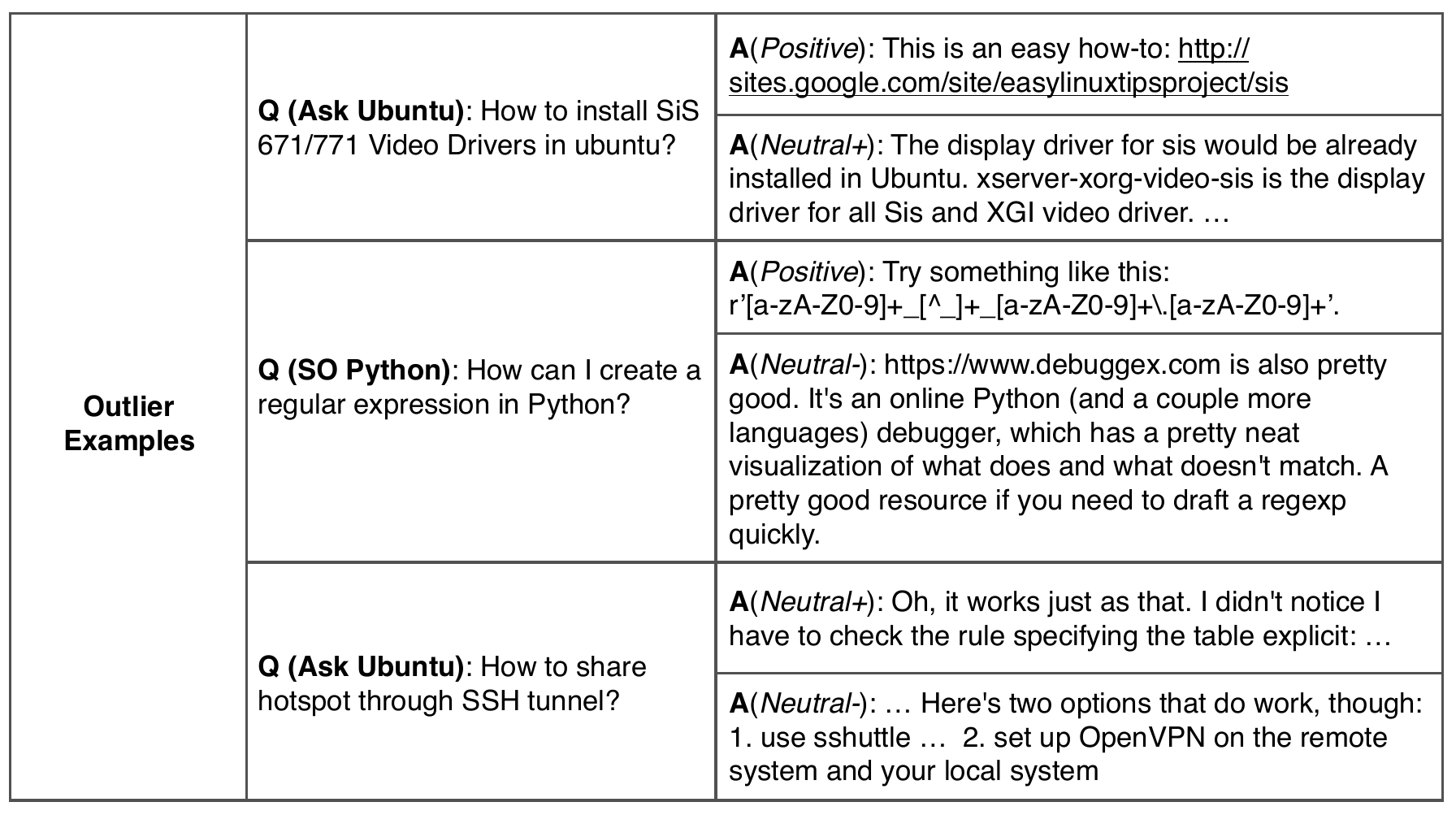}}
\caption{\revv{Outlier Examples in Label Establishment}}
\label{fig:outlier}
\end{figure}

\subsection{\revv{Outlier Cases Study}}
\revv{
As detailed in Section~\ref{subsec:label_establish}, we build our training samples via four heuristic rules, we thus can not ensure that there are no outlier cases distant from our heuristic rules. 
The outlier cases will produce a series of wrong preference pairs and hinder the learning performance of our model. 
Fig.~\ref{fig:outlier} shows three outlier examples for label establishment. From the figure we can see that:
\begin{enumerate}
    \item From the first example, it can be seen that, the quality of its non-accept answer in terms of informativeness and relevance are better than the accepted ones, not to mention that the link provided within the \emph{Positive} sample has been not available. This shows the outlier case that the non-accept answers may be better than the accepted answers. 
    \item From the second example, it can be seen that, for a given question, the answers from its similar questions are more descriptive than its own. This shows the outlier cases that the answers of other questions may be better than its own.
    \item From the last example, it can be seen that, the answers from its similar questions may provide more information cues than its non-accept answers.
\end{enumerate}
Detecting and removing these outlier cases before building the training samples will benefit the learning performance of our proposed {\sc DeepAns} model, we will focus on this research direction in the future.
}


%% file: related.tex
In this section, we describe the related studies on best answer retrieval, query expansion in software engineering, and deep learning in software engineering.

\subsection{Best Answer Retrieval}
Great effort has been dedicated to addressing the question answering tasks
on Q\&A sites~\cite{adamic2008knowledge, tian2013towards, calefato2019empirical, zhang2007expertise, jenders2016answer, calefato2016moving, sahu2016selecting, nie2017data}. 
Conventional techniques for retrieving answers primarily
focus on complementary features of the Q\&A sites.
For example, Adamic et al.~\cite{adamic2008knowledge} reported the first study on best answer prediction in Yahoo! Answers using user-related features. Following Adamic et al.'s study, Tian et al.~\cite{tian2013towards} trained a classifier on a dataset from Stack Overflow without relying on user-related features. Recently,  Calefato et al.~\cite{calefato2019empirical} modelled the answer prediction task as a binary classification problem, they assessed 26 best answer prediction model in Stack Overflow. Different from these works, we present a novel weakly supervised neural network architecture for ranking answers for a given question. To the best of our knowledge, our work is the first to apply deep neural network to the specific problem of best answer selection in Q\&A sites. Our approach can not only identify best answers from a list of candidate answers, but also recommend the most relevant answers for these unanswered posts. Besides, we also compare with Calefato et al.'s~\cite{calefato2019empirical} approach, and the experimental results have shown that the improvement is substantial. 

\subsection{Query Expansion in SE}  
Query expansion  has long been investigated as a way to improve the results returned by a search engine~\cite{haiduc2013automatic, hill2014nl, lu2015query, xu2017answerbot,rao-daume-iii-2018-learning, nie2016query, li2016query, azizan2015query, huang2018query}. Some software engineering researchers have employed query expansion to improve  the performance of tasks such as code search, answer summary, and similar question recommendation. For example, Haiduc et al.~\cite{haiduc2013automatic} proposed an approach that can recommend a good query reformulation strategy by performing machine learning on a set of historical queries and relevant
results. Hill et al.~\cite{hill2014nl} proposed a query expansion tool named Conquer,  which introduces a novel natural language based approach to organize and present search results and suggest alternative query words. 
More recently, Lu et at.~\cite{lu2015query} presented an approach to expand the original query with synonyms from WordNet, which can help developers to quickly reformulate a better query. Xu et al.~\cite{xu2017answerbot} proposed a novel framework to reformulate the answer in Stack Overflow to reduce the lexical gap between question and answer sentences. Inspired by these studies we also leverage the idea of query expansion to recommend the relevant answers.  Our \emph{DeepAns} tool generates useful clarifying questions as a way of query boosting, which can substantially reduce the lexical gap between the question and answer sentences. In contrast, all of the aforementioned studies  ignore the interactions between the asker and the potential helper.  

\subsection{Deep Learning in SE}
Recently, an interesting direction in software engineering is to use deep learning to solve many diverse software engineering tasks~\cite{white2019sorting, gu2018deep, hu2018deep, gao2020checking, li2017cclearner, huang2018automating, yang2015deep, gu2016deep, sun2019grammar, alahmadi2018accurately, kim2019guiding, kim2018generating, dong2019tablesense, chen2018neural, liu2017automatic, gao2019smartembed, gao2020code2que}. For example, 
White et al.~\cite{white2019sorting} leverage an deep learning approach, DeepRepair, for automatic program repairing.
Gu et al.~\cite{gu2018deep} propose a novel deep neural network named DeepCS for code search tasks, where code snippets semantically related to a query can be effectively retrieved. 
Hu et al.~\cite{hu2018deep} a develop new sequence-to-sequence model named DeepCom to automatically generate code comments for Java methods. 
Li et al.~\cite{li2017cclearner} present CClearner which is a deep-learning based approach for clone detection. 

Although the aforementioned studies have utilized deep learning techniques for different kinds of software engineering tasks, to our best knowledge, no one has yet considered the relevant answer recommendation  task in technical Q\&A sites. We proposed in this paper a novel neural network architecture to address the \revv{answer hungry} problems in  technical Q\&A forums.

%% file: conclusion.tex
To alleviate the \emph{\revv{answer hungry}} problem in technical Q\&A sites, we have presented a novel neural network-based tool, {\sc DeepAns}, to identify the most relevant answer among a set of answer candidates. Our model follows a three-stage process:\emph{question boosting}, \emph{label establishing} and \emph{answer recommendation}. 
Given a post, we first generate a clarifying question as a way of question boosting, 
we then automatically generate \emph{positive}, \emph{neutral$^+$}, \emph{neutral$^-$} and \emph{negative} training samples via label establishing. Finally based on the four kinds of training samples we generated, we trained a weakly-supervised neural network to compute the matching score between the question and candidate answers. 
\revv{Extensive experiments on the real-world technical Q\&A sites have comparatively demonstrated the promising performance and the robustness of our approach in solving unanswered/unresolved questions.}

%% file: acknowledgements.tex
This research was partially supported by the Australian Research Council’s Discovery Early Career Researcher Award (DECRA) funding scheme (DE200100021), ARC Laureate Fellowship funding scheme (FL190100035), ARC Discovery grant DP170101932 and Singapore’s Ministry of Education (MOE2019-T2-1-193).